\documentclass[%
 reprint,
superscriptaddress,
 amsmath,amssymb,
 aps,
]{revtex4-2}
\usepackage{graphicx}
\usepackage{longtable}
\usepackage{dcolumn}
\usepackage{bm}
\usepackage{hyperref}
\usepackage[mathlines]{lineno}

\begin{document}

\preprint{APS/123-QED}

\title{Differential cross sections for $\Lambda(1520)$ using photoproduction at CLAS\\}

\newcommand*{\ANL}{Argonne National Laboratory, Argonne, Illinois 60439}
\newcommand*{\ANLindex}{1}
\affiliation{\ANL}
\newcommand*{\ASU}{Arizona State University, Tempe, Arizona 85287-1504}
\newcommand*{\ASUindex}{2}
\affiliation{\ASU}
\newcommand*{\CANISIUS}{Canisius College, Buffalo, NY}
\newcommand*{\CANISIUSindex}{3}
\affiliation{\CANISIUS}
\newcommand*{\CMU}{Carnegie Mellon University, Pittsburgh, Pennsylvania 15213}
\newcommand*{\CMUindex}{4}
\affiliation{\CMU}
\newcommand*{\CUA}{Catholic University of America, Washington, D.C. 20064}
\newcommand*{\CUAindex}{5}
\affiliation{\CUA}
\newcommand*{\SACLAY}{IRFU, CEA, Universit\'{e} Paris-Saclay, F-91191 Gif-sur-Yvette, France}
\newcommand*{\SACLAYindex}{6}
\affiliation{\SACLAY}
\newcommand*{\CNU}{Christopher Newport University, Newport News, Virginia 23606}
\newcommand*{\CNUindex}{7}
\affiliation{\CNU}
\newcommand*{\UCONN}{University of Connecticut, Storrs, Connecticut 06269}
\newcommand*{\UCONNindex}{8}
\affiliation{\UCONN}
\newcommand*{\DUKE}{Duke University, Durham, North Carolina 27708-0305}
\newcommand*{\DUKEindex}{9}
\affiliation{\DUKE}
\newcommand*{\DUQUESNE}{Duquesne University, 600 Forbes Avenue, Pittsburgh, PA 15282 }
\newcommand*{\DUQUESNEindex}{10}
\affiliation{\DUQUESNE}
\newcommand*{\FU}{Fairfield University, Fairfield CT 06824}
\newcommand*{\FUindex}{11}
\affiliation{\FU}
\newcommand*{\FERRARAU}{Universita' di Ferrara , 44121 Ferrara, Italy}
\newcommand*{\FERRARAUindex}{12}
\affiliation{\FERRARAU}
\newcommand*{\FIU}{Florida International University, Miami, Florida 33199}
\newcommand*{\FIUindex}{13}
\affiliation{\FIU}
\newcommand*{\FSU}{Florida State University, Tallahassee, Florida 32306}
\newcommand*{\FSUindex}{14}
\affiliation{\FSU}
\newcommand*{\GWUI}{The George Washington University, Washington, DC 20052}
\newcommand*{\GWUIindex}{15}
\affiliation{\GWUI}
\newcommand*{\ISU}{Idaho State University, Pocatello, Idaho 83209}
\newcommand*{\ISUindex}{16}
\affiliation{\ISU}
\newcommand*{\INFNFE}{INFN, Sezione di Ferrara, 44100 Ferrara, Italy}
\newcommand*{\INFNFEindex}{17}
\affiliation{\INFNFE}
\newcommand*{\INFNFR}{INFN, Laboratori Nazionali di Frascati, 00044 Frascati, Italy}
\newcommand*{\INFNFRindex}{18}
\affiliation{\INFNFR}
\newcommand*{\INFNGE}{INFN, Sezione di Genova, 16146 Genova, Italy}
\newcommand*{\INFNGEindex}{19}
\affiliation{\INFNGE}
\newcommand*{\INFNRO}{INFN, Sezione di Roma Tor Vergata, 00133 Rome, Italy}
\newcommand*{\INFNROindex}{20}
\affiliation{\INFNRO}
\newcommand*{\INFNTUR}{INFN, Sezione di Torino, 10125 Torino, Italy}
\newcommand*{\INFNTURindex}{21}
\affiliation{\INFNTUR}
\newcommand*{\INFNPAV}{INFN, Sezione di Pavia, 27100 Pavia, Italy}
\newcommand*{\INFNPAVindex}{22}
\affiliation{\INFNPAV}
\newcommand*{\ORSAY}{Universit'{e} Paris-Saclay, CNRS/IN2P3, IJCLab, 91405 Orsay, France}
\newcommand*{\ORSAYindex}{23}
\affiliation{\ORSAY}
\newcommand*{\Juelich}{Institute fur Kernphysik (Juelich), Juelich, Germany}
\newcommand*{\Juelichindex}{24}
\affiliation{\Juelich}
\newcommand*{\JMU}{James Madison University, Harrisonburg, Virginia 22807}
\newcommand*{\JMUindex}{25}
\affiliation{\JMU}
\newcommand*{\KNU}{Kyungpook National University, Daegu 41566, Republic of Korea}
\newcommand*{\KNUindex}{26}
\affiliation{\KNU}
\newcommand*{\LAMAR}{Lamar University, 4400 MLK Blvd, PO Box 10046, Beaumont, Texas 77710}
\newcommand*{\LAMARindex}{27}
\affiliation{\LAMAR}
\newcommand*{\MISS}{Mississippi State University, Mississippi State, MS 39762-5167}
\newcommand*{\MISSindex}{28}
\affiliation{\MISS}
\newcommand*{\ITEP}{National Research Centre Kurchatov Institute - ITEP, Moscow, 117259, Russia}
\newcommand*{\ITEPindex}{29}
\affiliation{\ITEP}
\newcommand*{\UNH}{University of New Hampshire, Durham, New Hampshire 03824-3568}
\newcommand*{\UNHindex}{30}
\affiliation{\UNH}
\newcommand*{\NSU}{Norfolk State University, Norfolk, Virginia 23504}
\newcommand*{\NSUindex}{31}
\affiliation{\NSU}
\newcommand*{\OHIOU}{Ohio University, Athens, Ohio  45701}
\newcommand*{\OHIOUindex}{32}
\affiliation{\OHIOU}
\newcommand*{\ODU}{Old Dominion University, Norfolk, Virginia 23529}
\newcommand*{\ODUindex}{33}
\affiliation{\ODU}
\newcommand*{\JLU}{II Physikalisches Institut der Universitaet Giessen, 35392 Giessen, Germany}
\newcommand*{\JLUindex}{34}
\affiliation{\JLU}
\newcommand*{\PNU}{Pukyong National University, Busan 48513, Republic of Korea}
\newcommand*{\PNUindex}{35}
\affiliation{\PNU}
\newcommand*{\URICH}{University of Richmond, Richmond, Virginia 23173}
\newcommand*{\URICHindex}{36}
\affiliation{\URICH}
\newcommand*{\ROMAII}{Universita' di Roma Tor Vergata, 00133 Rome Italy}
\newcommand*{\ROMAIIindex}{37}
\affiliation{\ROMAII}
\newcommand*{\MSU}{Skobeltsyn Institute of Nuclear Physics, Lomonosov Moscow State University, 119234 Moscow, Russia}
\newcommand*{\MSUindex}{38}
\affiliation{\MSU}
\newcommand*{\SCAROLINA}{University of South Carolina, Columbia, South Carolina 29208}
\newcommand*{\SCAROLINAindex}{39}
\affiliation{\SCAROLINA}
\newcommand*{\TEMPLE}{Temple University,  Philadelphia, PA 19122 }
\newcommand*{\TEMPLEindex}{40}
\affiliation{\TEMPLE}
\newcommand*{\JLAB}{Thomas Jefferson National Accelerator Facility, Newport News, Virginia 23606}
\newcommand*{\JLABindex}{41}
\affiliation{\JLAB}
\newcommand*{\UTFSM}{Universidad T\'{e}cnica Federico Santa Mar\'{i}a, Casilla 110-V Valpara\'{i}so, Chile}
\newcommand*{\UTFSMindex}{42}
\affiliation{\UTFSM}
\newcommand*{\INSUBRIA}{Universit\`{a} degli Studi dell'Insubria, 22100 Como, Italy}
\newcommand*{\INSUBRIAindex}{43}
\affiliation{\INSUBRIA}
\newcommand*{\BRESCIA}{Universit\`{a} degli Studi di Brescia, 25123 Brescia, Italy}
\newcommand*{\BRESCIAindex}{44}
\affiliation{\BRESCIA}
\newcommand*{\GLASGOW}{University of Glasgow, Glasgow G12 8QQ, United Kingdom}
\newcommand*{\GLASGOWindex}{45}
\affiliation{\GLASGOW}
\newcommand*{\YORK}{University of York, York YO10 5DD, United Kingdom}
\newcommand*{\YORKindex}{46}
\affiliation{\YORK}
\newcommand*{\VIRGINIA}{University of Virginia, Charlottesville, Virginia 22901}
\newcommand*{\VIRGINIAindex}{47}
\affiliation{\VIRGINIA}
\newcommand*{\WM}{College of William and Mary, Williamsburg, Virginia 23187-8795}
\newcommand*{\WMindex}{48}
\affiliation{\WM}
\newcommand*{\YEREVAN}{Yerevan Physics Institute, 375036 Yerevan, Armenia}
\newcommand*{\YEREVANindex}{49}
\affiliation{\YEREVAN}

\newcommand*{\NOWHAMPTON}{Hampton University, Hampton, VA 23668}
\newcommand*{\NOWOHIOU}{Ohio University, Athens, Ohio  45701}
\newcommand*{\NOWISU}{Idaho State University, Pocatello, Idaho 83209}
\newcommand*{\NOWBRESCIA}{Universit\`{a} degli Studi di Brescia, 25123 Brescia, Italy}
\newcommand*{\NOWJLAB}{Thomas Jefferson National Accelerator Facility, Newport News, Virginia 23606}

\author {U.~Shrestha}
\email{Corresponding author: us810916@ohio.edu}
\affiliation{\OHIOU}
\author {T. Chetry}
\affiliation{\MISS}
\author {C.~Djalali} 
\affiliation{\OHIOU}
\author {K.~Hicks} 
\affiliation{\OHIOU}
\author {S.~i.~Nam}
\affiliation{\PNU}
\author {K.P. ~Adhikari} 
\altaffiliation[Current address:]{\NOWHAMPTON}
\affiliation{\ODU}
\author {S. Adhikari} 
\affiliation{\FIU}
\author {M.J.~Amaryan} 
\affiliation{\ODU}
\author {G.~Angelini} 
\affiliation{\GWUI}
\author {H.~Atac} 
\affiliation{\TEMPLE}
% ...

\collaboration{The CLAS Collaboration}
\noaffiliation

\date{\today}

\begin{abstract}
The reaction $\gamma p \rightarrow K^{+} \Lambda(1520)$ using photoproduction data from the CLAS $g12$ experiment at Jefferson Lab is studied. The decay of $\Lambda(1520)$ into two exclusive channels, $\Sigma^{+}\pi^{-}$ and $\Sigma^{-}\pi^{+}$, is studied from the detected $K^{+}$, $\pi^{+}$, and $\pi^{-}$ particles. A good agreement is established for the $\Lambda(1520)$ differential cross sections with the previous CLAS measurements. The differential cross sections as a function of CM angle are extended to higher photon energies. Newly added are the differential cross sections as a function of invariant 4-momentum transfer $t$, which is the natural variable to use for a theoretical model based on  a Regge-exchange reaction mechanism. No new $N^*$ resonances decaying into the $K^+\Lambda(1520)$ final state are found.
\end{abstract}

\maketitle

\section{\label{sec:introduction}Introduction\\}

Resonance structures are the signatures of excited valence quarks inside the nucleon. These excited resonances can then decay to a lower energy configuration by emitting a quark-antiquark pair. Hadron spectroscopy searches for these ground state and excited state baryons $(qqq)$, and their decay channels into mesons $(q\overline{q})$. The main objective is to identify the different quantum states (resonances) that come from analysis of their energies, widths, and characteristic line profiles. The study of baryon spectra is crucial to understanding Quantum Chromo-Dynamics (QCD).

Non-relativistic constituent quark models (NRCQMs) \cite{Isgur,IsgurKarl} can be considered as a naive and solvable approach to formulate hadronic wave functions in order to make predictions for the properties of baryonic ground states and excited states. They are, however, not so accurate at higher mass hadron spectra, when compared to experimental results~\cite{pdg2018}. Lattice QCD calculations \cite{latticeQCD} have shown ``missing resonances'' \cite{Bennhold1999} and other excited states, and are able to predict masses in the hadron spectra, but the currently available calculations are made at higher than physical masses because of the computational cost and therefore have limited accuracy. The systematic study of different decay channels is critical to the search for these missing resonances. Different studies have been carried out at Jefferson Lab. One approach has been to measure the strangeness photoproduction into two-body final states. Another one has shown the importance of three-body final states in order to study higher mass missing resonances \cite{mokeev2020}.

Photoproduction is an important mechanism to decipher information that identifies the dynamical basis behind ground state and excited state resonance structure formation. This study aims to deliver a better understanding of photoproduction of the hyperon resonance, $\Lambda(1520)$.

The framework of this paper is the following. Section~\ref{sec:L1520studies} presents a brief summary of previous experimental and theoretical studies on the photoproduction of the $\Lambda(1520)$ hyperon. Section~\ref{sec:expsetup} introduces the experimental set up that provided the data for this study. Section~\ref{sec:data} outlines the details of the event selection, simulation, and yield extraction procedures for cross sections in Sections~\ref{sec:eventselection},~\ref{sec:simulation}, and~\ref{sec:yieldandacceptance}. The measured cross sections are displayed in Section~\ref{sec:diffcross}. Section~\ref{sec:sysunc} gives an account of the systematic uncertainties for this study. The comparison of the results with the theoretical predictions are discussed and our conclusions are provided in Section~\ref{sec:discussion}.

\section{\label{sec:L1520studies}Previous Studies on $\Lambda(1520)$\\}

The $\Lambda(1520)$ has been well studied \cite{pdgLambda2018}. The Laser Electron Photon Experiment at SPring (LEPS) Collaboration studied the photoproduction of $\Lambda(1520)$ and measured differential cross sections and photon-beam asymmetries with linearly polarized photon beams in the energy range $1.5 < E_{\gamma} < 2.4$~GeV at forward angles \cite{LEPS2010}, where the authors reported a bump structure at $W \simeq 2.11$~GeV suggesting a nucleon resonance or a new reaction process. Another photoproduction study \cite{LEPS2009} for this hyperon by LEPS with liquid hydrogen and deuterium targets at similar photon energies, showed a reduced production from neutrons compared to protons at backward angles.

A recent study on $\Lambda(1520)$ photoproduction was completed by Moriya \textit{et al}.\ \cite{kmoriya}. As part of the $g11$ experiment, this was done at the CEBAF Large Accepteance Spectrometer (CLAS) with an unpolarized real photon beam at energies up to 4.0~GeV striking a liquid hydrogen target. 
They studied all of the three $\Sigma\pi$ decay modes, measuring the kaon angular distributions, which were flat at threshold ($W \simeq 2.0$~GeV) and forward peaked at energies above the threshold \cite{kmoriya}. A comparison of their results with results from this study is provided in Section~\ref{sec:diffcross}. Even though Moriya's result compared well with the model predictions of He and Chen \cite{He2012} and Nam \textit{et al}.\ \cite{Nam2010} for the $\Lambda(1520)$, our results will extend the cross sections to higher photon energies using the $g12$ experiment at CLAS, where the current theoretical models of Regge exchange are expected to be more accurate \cite{Nam2010}.

The photoproduction of $\Lambda(1520)$ off a proton target has been studied theoretically by Nam \textit{et al}.\ \cite{Nam2010}. The authors investigated the $\Lambda(1520)3/2^-$, or $\Lambda^*$, photoproduction in an effective-Lagrangian approach using Born terms where they use the Rarita-Schwinger formalism to account for the spin-3/2 fermion field. They introduced hadron form factors that represent spatial distributions for hadrons, and in order to preserve gauge invariance of the invariant amplitude, they also include a contact term \cite{Haberzettl1998}. One of several free parameters in their model, the vector-kaon coupling constant $g_{K^*N\Lambda^*}$, has been constrained by data, along with the anomalous magnetic moment of the $\Lambda^*$, $\kappa_{\Lambda^*}$, for photon energies below 3~GeV.

Regge theory accounts for the analytic properties of scattering as a function of angular momentum. The theory uses Regge trajectory functions $\alpha(s$, $t)$, where $s$ and $t$ are Mandelstam variables, that can correlate certain sequences of particles or resonances. Regge theory accounts for the exchange of entire families of hadrons, with identical internal quantum numbers but different spins $J$. In order to extend their model to higher energy, Nam \textit{et al}.\ \cite{Nam2010} have implemented the Regge contributions in $\Lambda^*$ photoproduction by considering mesonic Regge trajectories, corresponding to all the meson exchanges with the same quantum numbers but different spins in the $t$ channel at tree level.

He and Chen \cite{He2012} also studied an effective-Lagrangian approach model for the $\Lambda(1520)$, which takes into consideration the vector meson $K^*$ exchanged in the $t$ channel, which has proven to be significant at high energy (11~GeV). Besides the Born terms, the inclusion of the contact term and $s$, $u$, and $K$ exchanged $t$ channels have important contribution at all energies. They report a contribution from the nucleon resonance $N(2080)3/2^-$ in $\Lambda(1520)$ photoproduction that suggests the need for another resonance at a nearby mass.
Studies of $\Lambda^*$ photoproduction help to strengthen the idea that the effective-Lagrangian approach is a valid theoretical model. Moriya \textit{et al}.\ \cite{kmoriya} compared their cross sections with the model calculation from Nam \textit{et al}.\ \cite{Nam2010} and He \textit{et al}.\ \cite{He2012}, and concluded that the latter model, because of the additional interaction ingredients, had a slightly better agreement with their data.

\begin{figure}
  \centering
  \includegraphics[width =  0.7\linewidth, keepaspectratio = true]{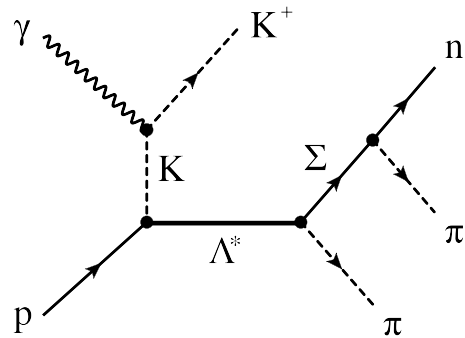}
  \caption{Schematic diagram for the reaction, $\gamma p \rightarrow K^+ \Lambda^*$, assuming $t$-channel dominance. $\Lambda^*$ here represents $\Lambda(1520)$.}
  \label{fig:reaction}
\end{figure}

Regge exchange ($t$-channel) models are expected to be more accurate as the beam energy goes above the resonance region.  For $W>3.0$~GeV, this is beyond the center of mass energy range where individual resonance contributions (in the $s$-channel) are significant, and hence the $g12$ data will provide a more stringent test of the model of Nam \textit{et al}.\ \cite{Nam2010}.

\subsection{$\Lambda(1520)$ Photoproduction}

Photoproduction off a proton can create a $K^{+}$-meson and a $\Lambda^*$ resonance, $\Lambda(1520)$, see Fig.~\ref{fig:reaction}, which can decay via $\Sigma\pi$ channels, \textit{e.g.}, $\Sigma^{+}\pi^{-}$, $\Sigma^{-}\pi^{+}$, and $\Sigma^{0}\pi^{0}$. For the two charged decay channels, $\Sigma^{\pm}$ gives off a neutron and a $\pi^{\pm}$. Both the $\Sigma^{+}$ and $\Sigma^{-}$ branches of the photoproduced $\Lambda(1520)$ have $K^{+}$, $\pi^{+}$, and $\pi^{-}$ as the detected final state particles.

\section{\label{sec:expsetup}Experimental Setup\\}

\subsection{The CLAS Detector}
The data set used in this analysis of $\Lambda(1520)$ photoproduction is taken from the $g12$ experiment performed by the CLAS Collaboration at the Thomas Jefferson National Accelerator Facility (TJNAF). Located in Newport News, Virginia, Jefferson Lab houses four experimental halls, namely, A, B, C, D, and the Continuous Electron Beam Accelerator Facility (CEBAF). The CLAS detector was installed inside the Hall~B and was decommissioned in 2012.

The design of the CLAS detector was based on a toroidal magnetic field that had the ability to measure charged particles with good momentum resolution, provide geometrical coverage of charged particles over a large laboratory angular range, and keep a magnetic field-free region around the target to allow the use of dynamically polarized targets \cite{clas}. Six superconducting coils around the beamline produced a field in the azimuthal direction. Drift chambers (DC) for charged particle trajectories, gas Cherenkov counters for electron identification, scintillation counters for time-of-flight (TOF), and electromagnetic calorimeters (EC) to detect showering electrons, photons, and neutrons, were the major components of the detector's particle identification (PID) system \cite{clas}.

The accelerated electron beam produced bremsstrahlung photons when passed through a thin gold radiator. The photon tagger system at CLAS used a hodoscope with two scintillator planes (energy and timing counters) that enabled photon tagging by the detection of energy-degraded electrons, which were deflected in the tagger magnetic field \cite{clas}. The start counter, surrounding the target, recorded the start time of the outgoing particles that originated in the target.    

\subsection{$g12$-Run}
The data set used in this analysis was acquired from the CLAS $g12$ experiment that was performed in the summer of 2008. This CLAS experiment was a
high-luminosity, high-energy, real-photon run. This run used an electron beam current of 60-65 nA that produced bremsstrahlung photons. The photons continued forward towards the
40 cm LH$_{2}$ (liquid hydrogen) target. The photon energy for the run was up to 5.7~GeV. Details of the $g12$ experiment, the running conditions, and the formulated standard procedures for the data analysis can be found on \textit{The g12 Analysis Procedure, Statistics, and Systematics} document \cite{g12note}.

\section{\label{sec:data}Data Analysis\\}
For the reaction $\gamma p \rightarrow K^+ \Lambda(1520) \rightarrow K^+ \Sigma^{\pm}\pi^{\mp}$, two exclusive decay channels, $\Lambda(1520) \rightarrow \Sigma^{\pm}\pi^{\mp} \rightarrow n \pi^{\pm}\pi^{\mp}$, were identified by detecting $K^+$, $\pi^+$, and $\pi^-$. The unmeasured $\Sigma^{\pm}$ and $n$ were reconstructed from the missing mass $(MM)$ approach. 

\subsection{\label{sec:eventselection}Event Selection}

\subsubsection{Photon Selection}
\label{sec:photonselection}
For the $g12$ experiment, the accelerator delivered electrons in packets of 2-ns bunches into Hall~B, where bremsstrahlung photons were then produced. A reaction inside the target was triggered by an incident photon. An event was recorded when a triggered reaction was associated with a specific photon candidate. Since there were several potential photon contenders for a recorded event due to background sources, it was necessary to determine the correct photon that created a specific event. 

In order to find the correct photon, the \textit{coincidence time}, $\Delta t_{coinc}$, was defined per photon as the difference between the Tagger time ($t_\gamma$) and the Start Counter time ($t_{event}$) extrapolated to the interaction point,
\begin{equation}
    \Delta t_{coinc} = t_{event} - t_{\gamma}.
\end{equation}
The Tagger time, also known as the photon time, is the time at which a photon reached at the point of interaction or the event vertex point, whereas the Start Counter time, also known as the event time, is understood as the average of the time per track of the particle, when detected by the Start Counter, at the same vertex point.

\begin{figure}
  \includegraphics[width=1\linewidth, keepaspectratio = true]{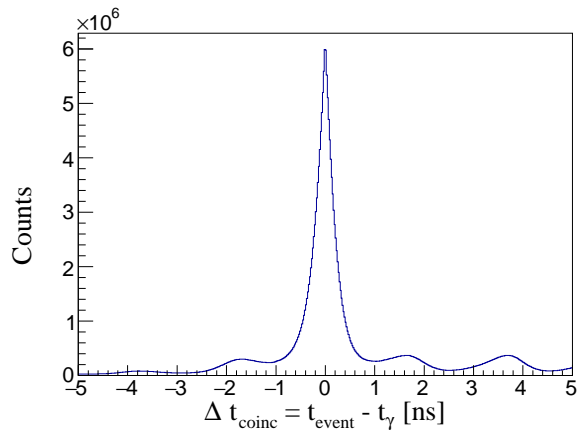}
  \caption{\label{fig:tbidcutdata}The photon coincident-time distribution, $\Delta t_{coinc}$, for the events with $K^{+}$, $\pi^{+}$, and $\pi^{-}$ as the detected particles is shown. The 2-ns bunching of the photon beam is apparent. Events with $\Delta t_{coinc} = |t_{event} - t_{\gamma}| < 1$~ns cut on the coincidence-time distribution are selected.}
\end{figure}

Figure~\ref{fig:tbidcutdata} shows a distribution of the coincidence time, where multiple photon candidates per event can be seen. A photon peak selection cut of $\Delta t_{coinc}=|t_{event}-t_{\gamma}|<1$~ns was applied.

There can be events with more than one photon with $|\Delta t_{coinc}| < 1$~ns. This is known as photon multiplicity \cite{g12note}. A correction factor, $\gamma_{corr} =\nolinebreak 1.03$, was obtained by examining photon multiplicities in both data and simulation, and was applied in the calculation of the differential cross sections.

\subsubsection{\label{sec:PID}Particle Identification}
\begin{figure*}
  \includegraphics[width=0.32\textwidth, keepaspectratio = true]{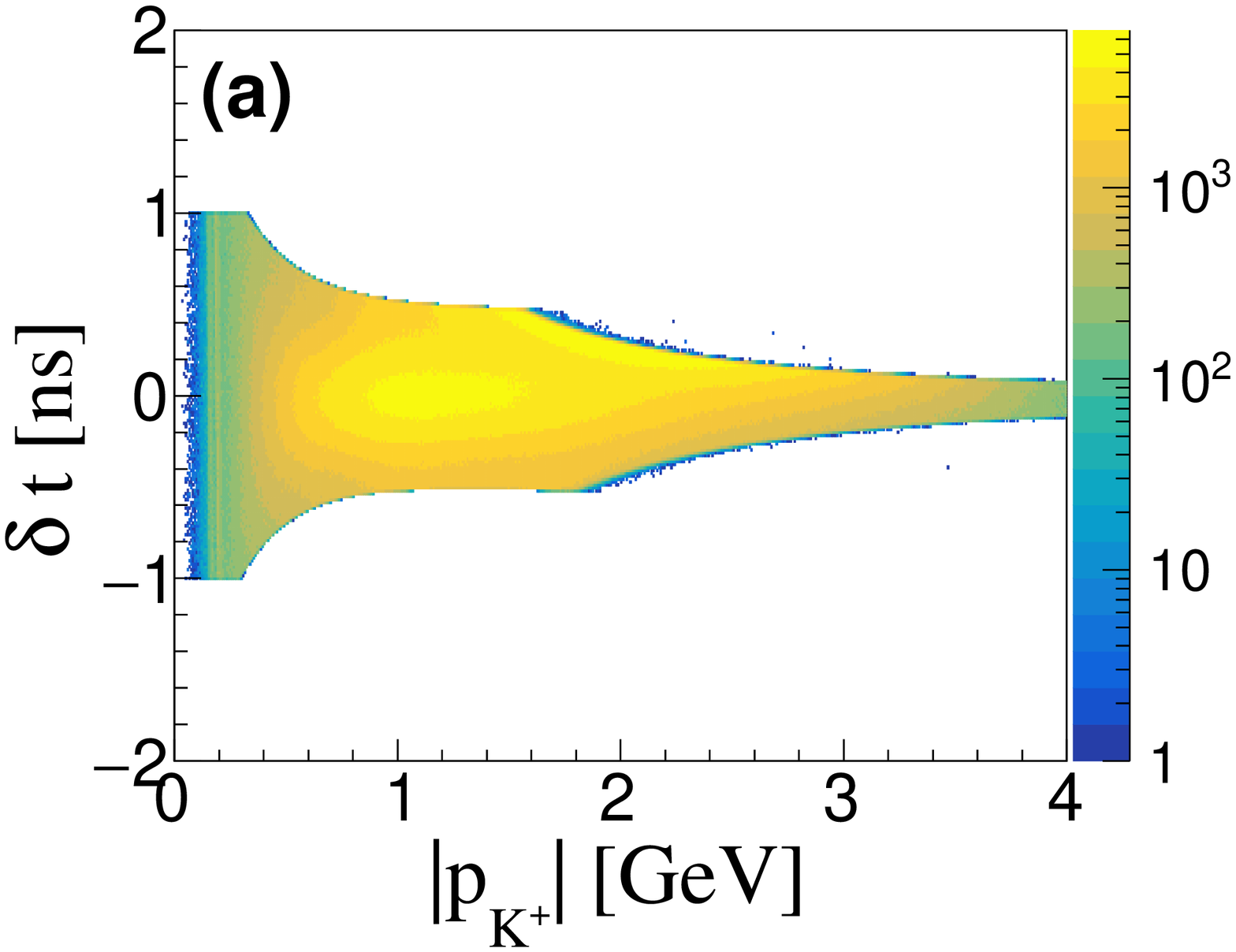}
  \includegraphics[width=0.32\textwidth, keepaspectratio = true]{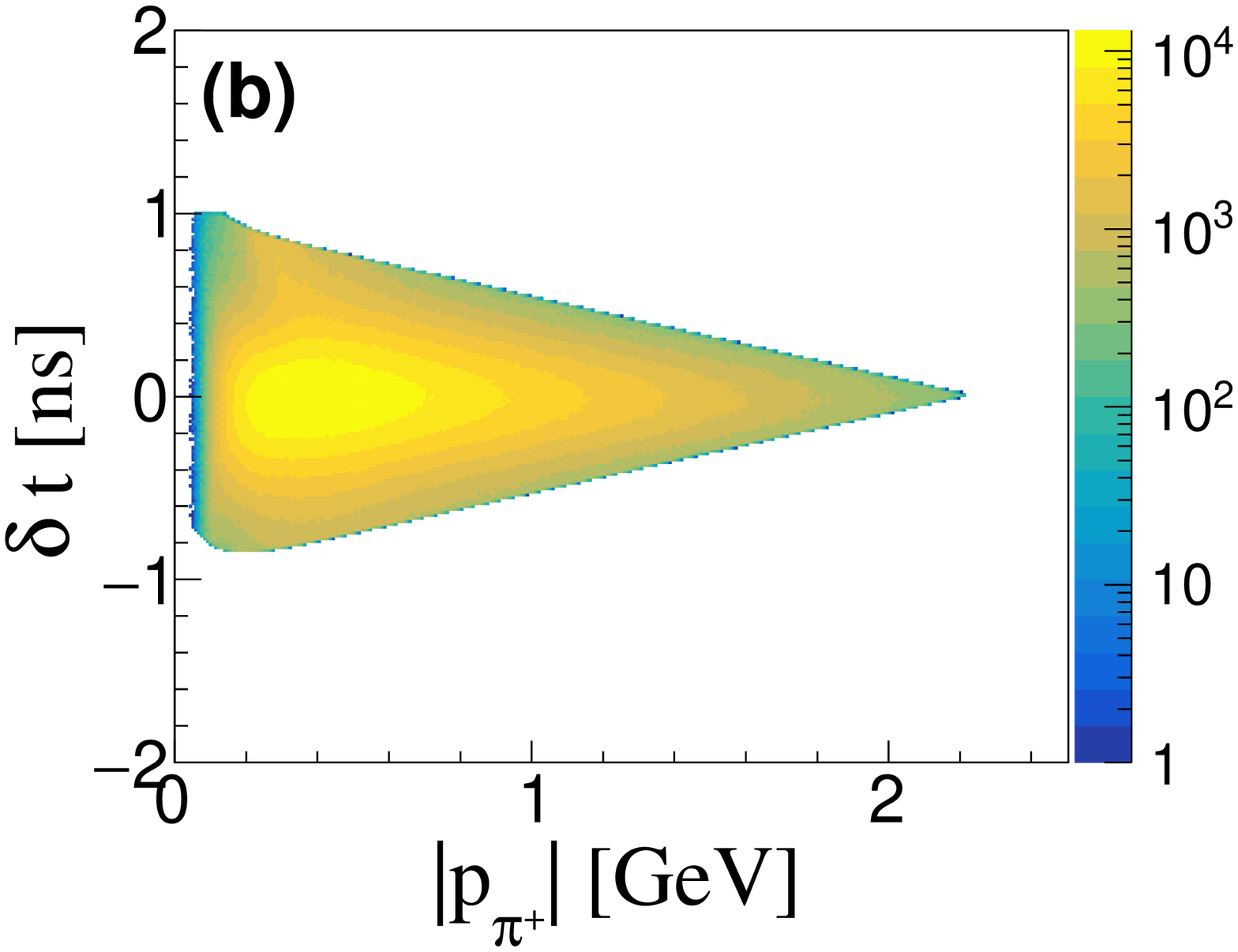}
  \includegraphics[width=0.32\textwidth, keepaspectratio = true]{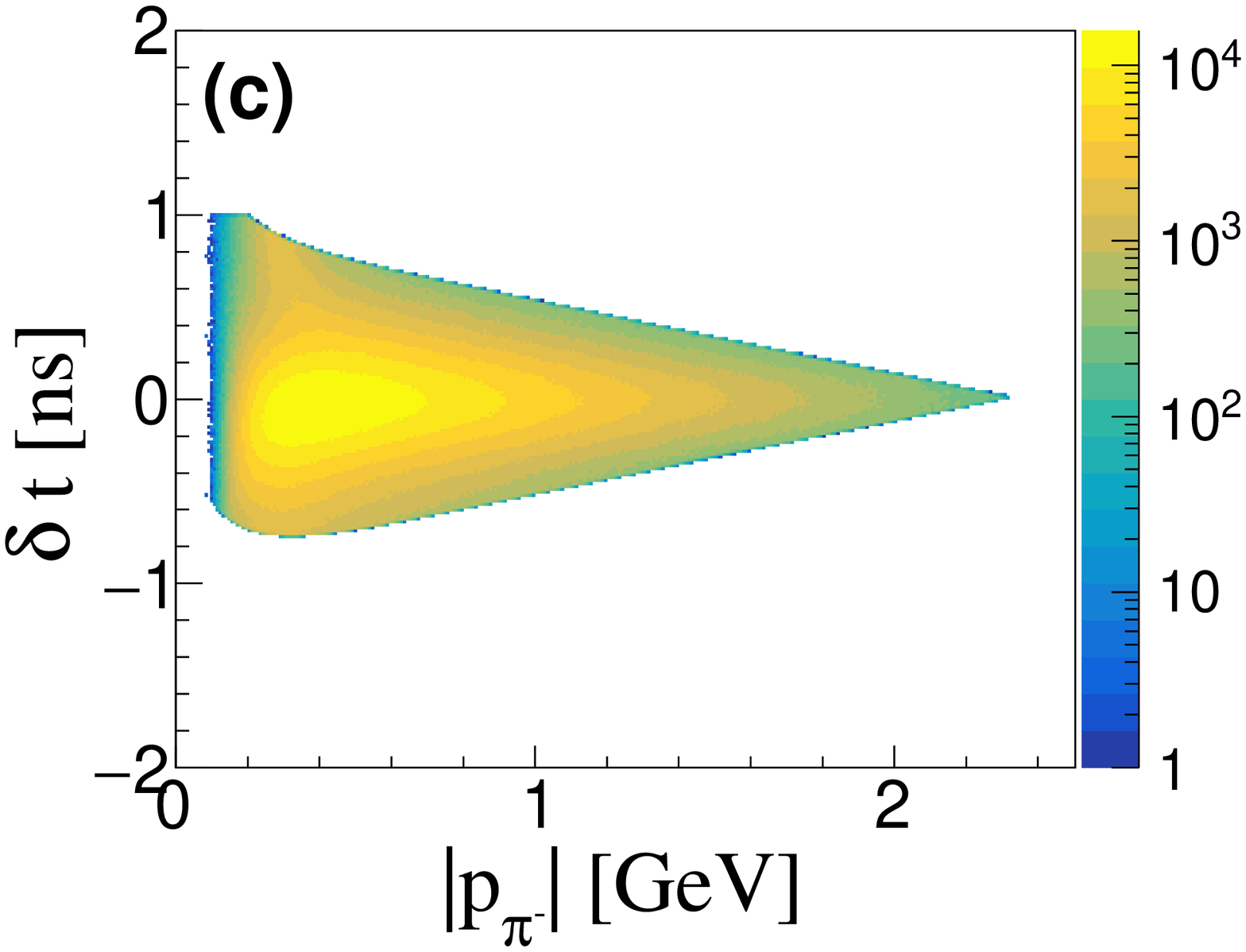}
  \caption{\label{fig:pidplotdata}The timing versus momentum distributions used for particle identification for $K^+$, $\pi^+$, and $\pi^-$ in (a), (b), and (c), respectively, for the data as a function of the particle's momentum are shown. Straight cuts of $|\delta t| < 1$~ns for each particle, along with a $\pm 2.5\sigma$ momentum-dependent timing cut around the centroid of $\delta t$ in each momentum bin are employed in identifying the particles coincident with a single photon.}
\end{figure*}

The particle identification (PID) process used the information signature left by a particle when it hit a detector
in order to identify that particle. Proper particle identification was vital to reduce backgrounds and improve measurement resolution. The PID information was accessed and the data were passed through the skimming process
where events that included the topologies with the final state particles were selected. PID process was refined by employing the time-of-flight technique to identify
particles. The measured time of flight ($t_{meas}$) for any particle produced during an event was compared with
the calculated time of flight ($t_{calc}$) for the known path ($d_{path}$) for the measured momentum ($p$) and an assumed mass ($m$).
The measured and calculated time of flight information used the measured and calculated $\beta$ values, $\beta_{meas}$, and $\beta_{calc}$, respectively.
The $\beta_{meas}$ was obtained from the CLAS-measured momentum $p$ of a particle during the data skimming process, whereas $\beta_{calc}$ is a theoretical value
calculated from that measured momentum and the particle's assumed mass. The time difference ($\delta t$) between the measured time ($t_{meas}$) and the calculated time ($t_{calc}$) is given by,
\begin{equation}
\delta t = t_{meas} - \frac{d_{path}}{c}\frac{E}{p} = \frac{d_{path}}{c} \left(\frac{1}{\beta_{meas}} - \frac{1}{\beta_{calc}}\right),
\end{equation}
where $E = \sqrt{p^2 + m^2}$ is the total energy of a particle. The $\delta t$ is recalculated for each particle based on its mass and charge.

For the first-order PID, the three detected particles, $K^+$, $\pi^+$, and $\pi^-$, were selected with a timing cut of $|\delta t| <\nolinebreak 1$~ns in the timing versus momentum distribution for each particle. More stringent momentum-dependent timing cuts were applied by binning the timing difference distributions for the charged particles peaking at $\delta t = 0$~ns, into several momentum bins and then fitting with a Gaussian function for $\pi^{\pm}$ or a Gaussian function over an exponential background for $K^+$. The extracted centroid and width parameters were used to apply a $\pm 2.5\sigma$ timing cut for both data and simulations to the original 1-ns timing-momentum distribution of the detected particles. Figure~\ref{fig:pidplotdata} shows the timing versus momentum distributions of the three detected particle after the PID cuts.

\subsubsection{\label{sec:othercuts}Minimum $|p|$, $z$-vertex, Fiducial, and Paddle Cuts}

The detection efficiency of low momentum particles was not large and not so accurately known. Such particles were eliminated by applying minimum momentum cuts $|p_{K^{+}}| < 0.35$~GeV, $|p_{\pi^{+}}| < 0.15$~GeV, and $|p_{\pi^{-}}| < 0.17$~GeV, where $|p|$ represents the magnitude of the particle momentum.

The $g12$ experiment used a liquid hydrogen (LH$_{2}$) target measuring 40 cm in length and 2 cm in diameter. The target was not centered around the CLAS detector, $z=0$~cm, but was shifted 90~cm upstream in order to increase the detector acceptance in the forward direction. Charged tracks from the target were selected in reconstruction by requiring that they came from the $z$ range (coordinate along the beamline) from -100 to -70~cm.

The geometry of the CLAS detector and the presence of the toroidal magnetic field could cause inaccurate reconstruction of particle tracks at the edges of the drift chambers, thereby resulting in uncertainties in the detector efficiency. We introduced fiducial boundaries that encompass a well-behaved and predictable acceptance region in azimuthal angle, $\phi$, of the detector for each particle depending upon its momentum, charge, and polar angle, $\theta$. Hence, a standard geometric fiducial cut procedure \cite{g12note} was applied for each detected particle in all six sectors, both for the data and for the simulation. 

Scintillation counters were used to determine the time of flight of charged particles. Counter with very low photomultiplier gain resulted in poor timing resolution and poor efficiency. These bad paddles were identified and removed from the data analysis during the event selection process \cite{g12note}.

\subsubsection{\label{sec:missingmasscuts}Missing Mass Cuts}

\begin{figure}
   \includegraphics[width= 1\linewidth, keepaspectratio = true]{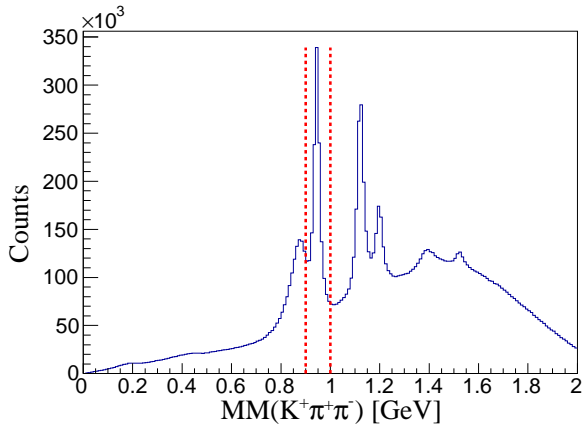}
    \caption{The missing mass distribution used to reconstruct the missing $n$ is shown. A $0.9\leq MM(K^+\pi\pi)$[GeV] $ \leq1.0$ cut to select neutrons is indicated by the dotted lines.}
  \label{fig:MM}
\end{figure}
A series of \textit{missing mass} cuts was applied to isolate and filter events corresponding to the two topologies for the $\Lambda(1520)$. In both of its charged decay channels, $\Lambda(1520)$ branches into $\pi^+$, $\pi^-$, and $n$. Since, $\pi^+$ and $\pi^-$ are the detected particles, the missing mass distribution given by,
\begin{equation}
    MM(K^+\pi^+\pi^-) = \sqrt{(P_{\gamma}+P_{p}-P_{K^+}-P_{\pi^+}-P_{\pi^-})^2},
\end{equation}
was constructed to select the missing $n$, where $P_{\gamma}$, $P_{p}$, $P_{K^+}$, $P_{\pi^+}$, and $P_{\pi^-}$ are the four-momenta of the incoming photon, target proton, and the outgoing particles, $K^+$, $\pi^+$, and $\pi^-$, respectively. The missing $n$ peak can be seen in Fig.~\ref{fig:MM} and the corresponding events with a $n$ were selected by making the cuts indicated in the figure.

A small structure seen at around 0.85~GeV to the left of the $n$ peak in Fig.~\ref{fig:MM} is due to $\pi^+$ tracks incorrectly reconstructed as $K^+$ tracks. Events from the three pion reaction, $\gamma p \rightarrow \pi^{+} \pi^{+} \pi^{-} n$, can have one of the $\pi^{+}$ misidentified as a $K^{+}$. These events form a nearly uniform background, and do not contribute to the yield of the $\Lambda(1520)$ as shown below. The side structure at 0.85~GeV relative to the $n$ peak is reduced in size by applying more stringent cuts in the particle identification process.

\begin{figure}
    \includegraphics[width= 1\linewidth, keepaspectratio = true]{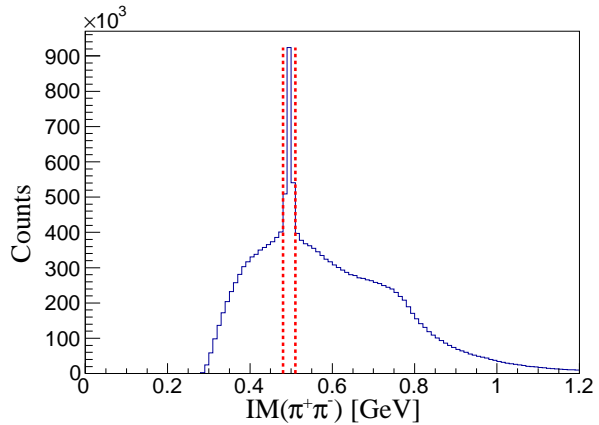}
  \caption{Removing $K^0\rightarrow\pi^+\pi^-$ ($\Lambda(1520) \rightarrow nK^0$ channel) by cutting out events with $0.48\leq IM(\pi^{+}\pi^{-})$[GeV] $ \leq0.51$ as indicated by the dotted lines.}
  \label{fig:impippim}
\end{figure}

It is important to consider the $nK^0$ decay of $\Lambda(1520)$. The $K^0$ can decay to a $\pi^+\pi^-$ pair. Hence, the $n$-cut
selected events mentioned earlier can have charged pions contribution to the final state particles. In order to exclusively look
for the $\Lambda(1520)$ photoproduction from the $\Sigma^{\pm}\pi^{\mp}$ channel, events from a possible $nK^0$ channel were excluded by removing events in the $K^0$ peak in the invariant mass distribution plot for $\pi^+\pi^-$, given by,
\begin{equation}
    IM(\pi^+\pi^-) = \sqrt{(P_{\pi^+} + P_{\pi^-})^2},
\end{equation}
as seen in Fig.~\ref{fig:impippim}.

\begin{figure*}
    \includegraphics[width= 0.49\linewidth, keepaspectratio = true]{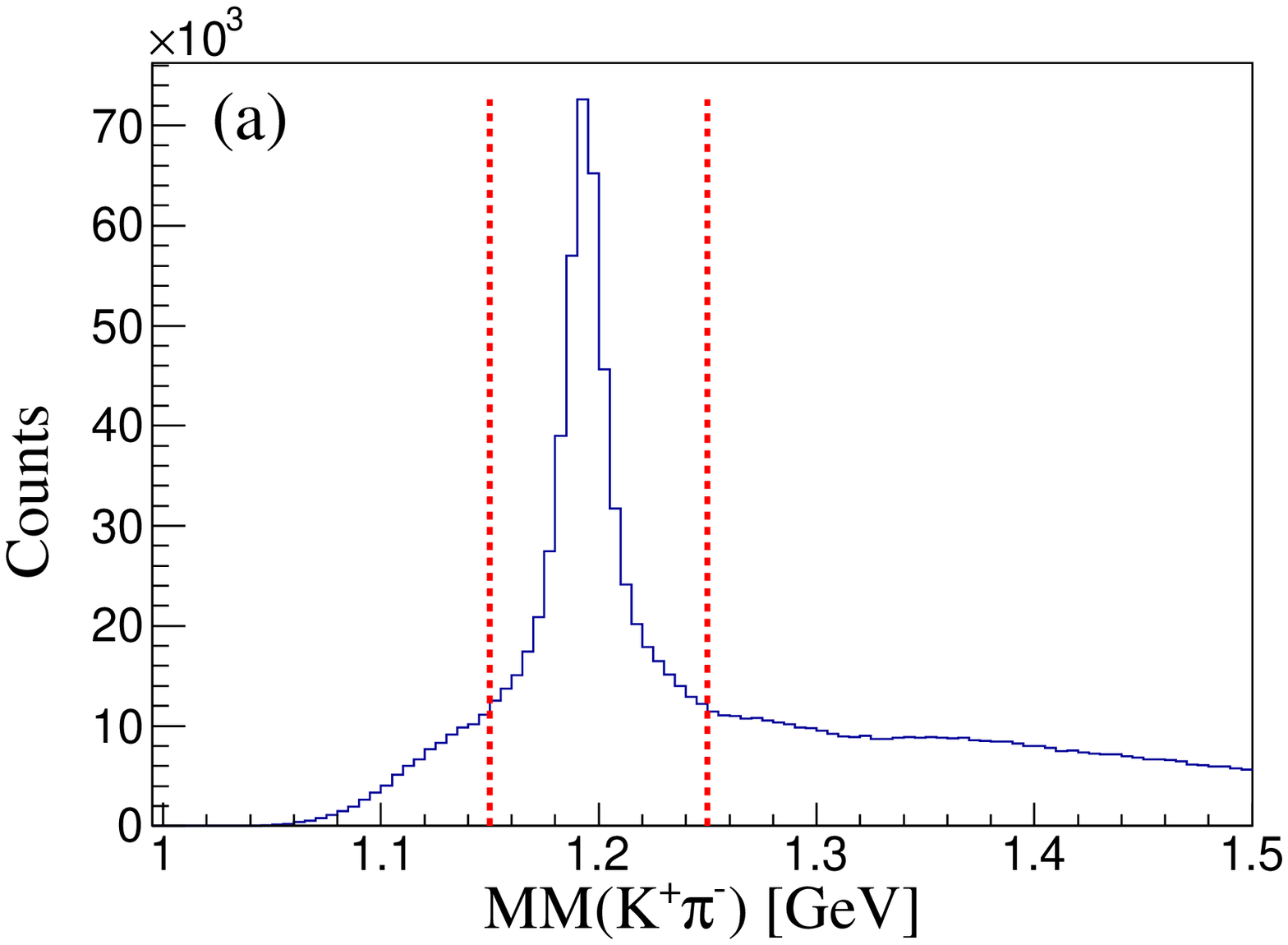}
    \includegraphics[width= 0.49\linewidth, keepaspectratio = true]{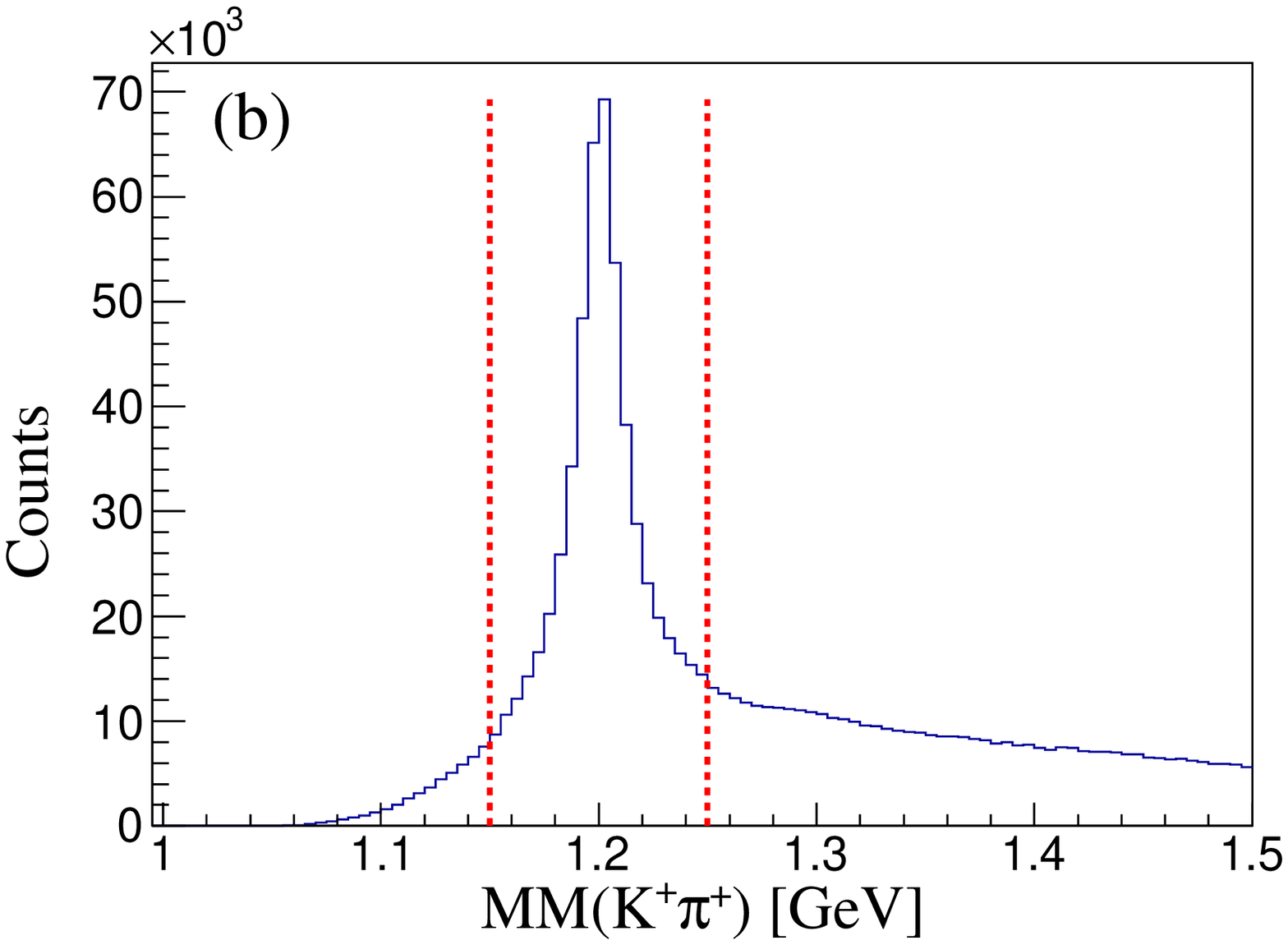}
  \caption{Selection of the $\Sigma^+$ (a) and $\Sigma^-$ (b) in the event distributions by making cuts $1.15 \leq MM(K^+\pi^-)$~[GeV] $\leq1.25$ and $1.15 \leq MM(K^+\pi^+)$~[GeV] $\leq1.25$, respectively. The cuts are shown by the dotted lines. The two decay branches are analyzed separately for further analysis.}
  \label{fig:sigpsigm}
\end{figure*}

After applying the above-mentioned cuts, the analysis branches into the two exclusive reaction channels for the $\Lambda(1520)$, $\Lambda(1520) \rightarrow \Sigma^+\pi^-$ and $\Lambda(1520) \rightarrow \Sigma^-\pi^+$. This was done by by plotting the missing mass distributions, $MM(K^+\pi^-)$ and $MM(K^-\pi^+)$ for the $\Sigma^+$ and $\Sigma^-$ channels, respectively. Straight cuts were applied to select events with $\Sigma^+$ and $\Sigma^-$ in their respective missing mass distributions, as shown in Fig.~\ref{fig:sigpsigm}. Even though the $\Sigma^+$ and $\Sigma^-$ event distributions in the data show some background, the background will not form a peak at the $\Lambda(1520)$. Some of the background in the data comes from generic background processes and reactions, such as $\gamma p \rightarrow K^{*0} \Sigma^{+}$ ($K^{*0} \rightarrow K^+ \pi^-$) and $\gamma p \rightarrow \pi^{+} \pi^{+} \pi^{-} n$ (where $\pi^{+}$ is misidentified as $K^{+}$). The latter reaction makes a smooth background under the $\Sigma$ peaks.

\begin{figure*}
    \includegraphics[width= 0.49\linewidth, keepaspectratio = true]{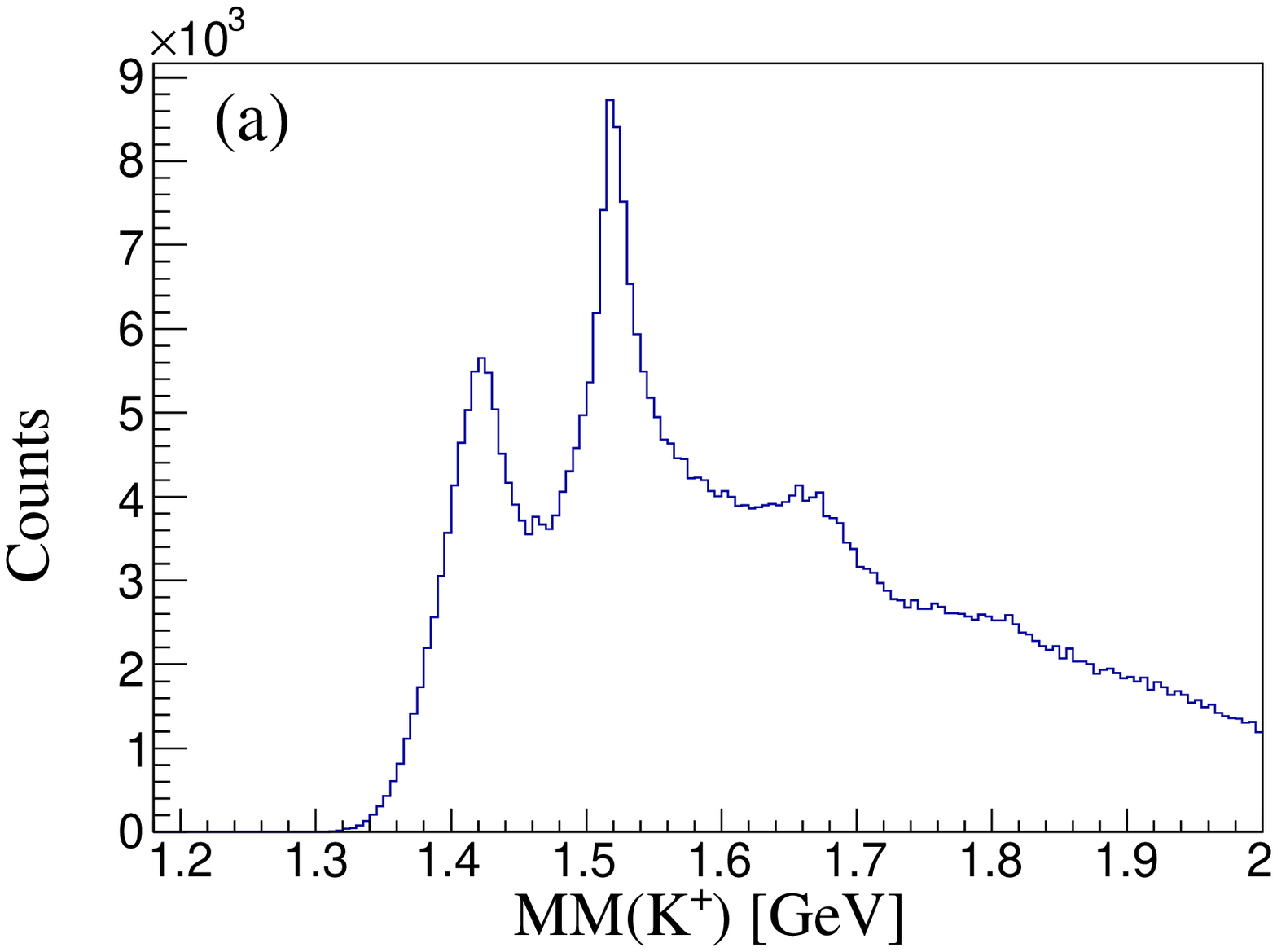}
    \includegraphics[width= 0.49\linewidth, keepaspectratio = true]{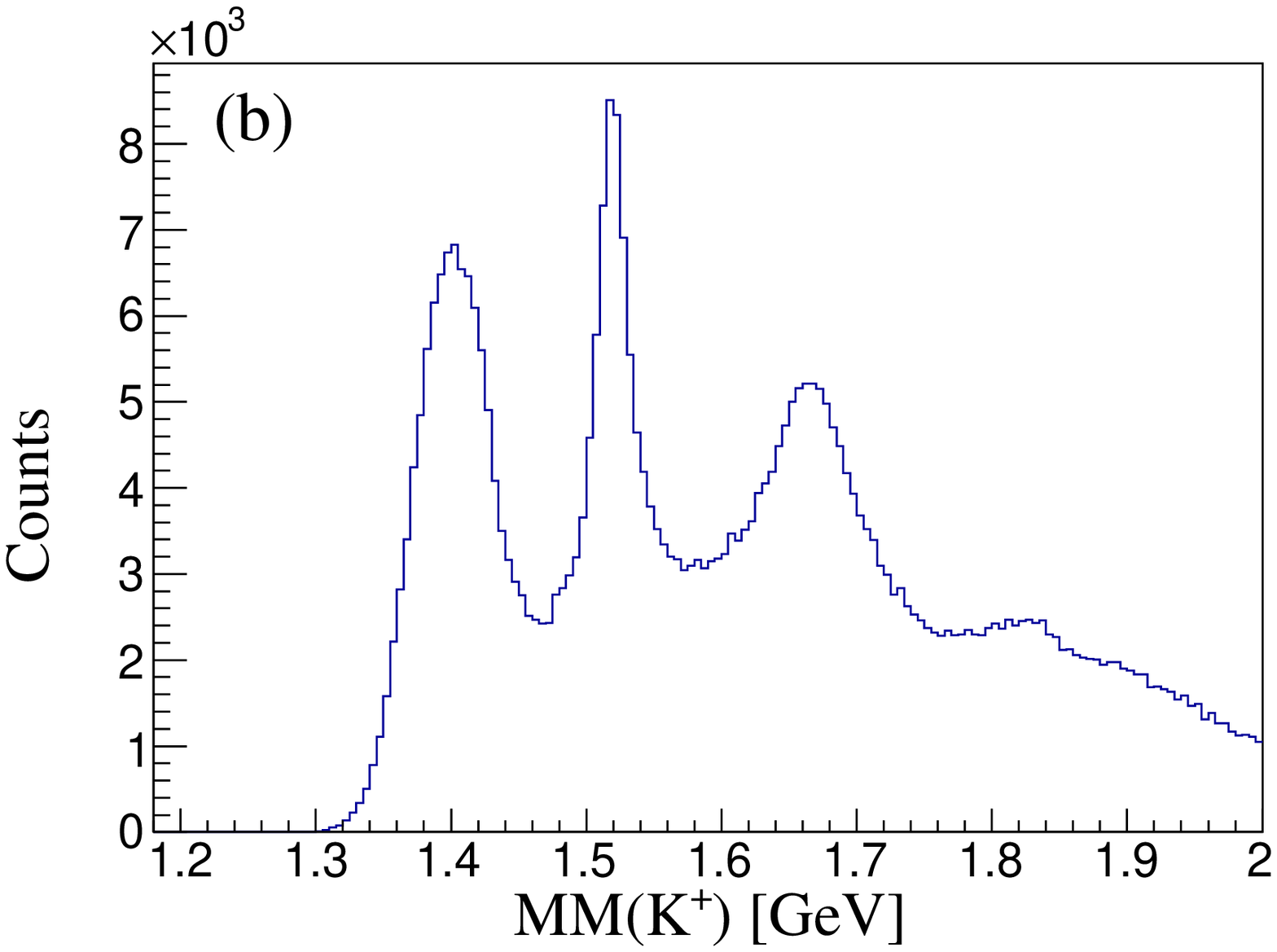}
  \caption{$\Lambda^*$ resonances for the $\Sigma^+$ and $\Sigma^-$ channels are shown in (a) and (b), respectively. The data distributions show the $\Lambda^*$ resonances $\Lambda(1405)$ and $\Lambda(1520)$. The wider peak around 1.68~GeV is due to higher-mass resonances $\Lambda(1670)$ and $\Lambda(1690)$.}
  \label{fig:alllambdasigpsigm}
\end{figure*}

Now the $\Lambda^*$ resonances can be seen in the $MM(K^+)$ distributions for the two channels as shown in Fig.~\ref{fig:alllambdasigpsigm}. The peaks at around 1.40~GeV and 1.52~GeV correspond to the $\Lambda(1405)$ and $\Lambda(1520)$. The smooth peak near 1.68~GeV represents the two higher mass resonances
$\Lambda(1670)$ and $\Lambda(1690)$. The analysis of those flavor-octet resonances is being carried out and will be the subject of a future publication.

\begin{figure}
   \includegraphics[width= 1\linewidth, keepaspectratio = true]{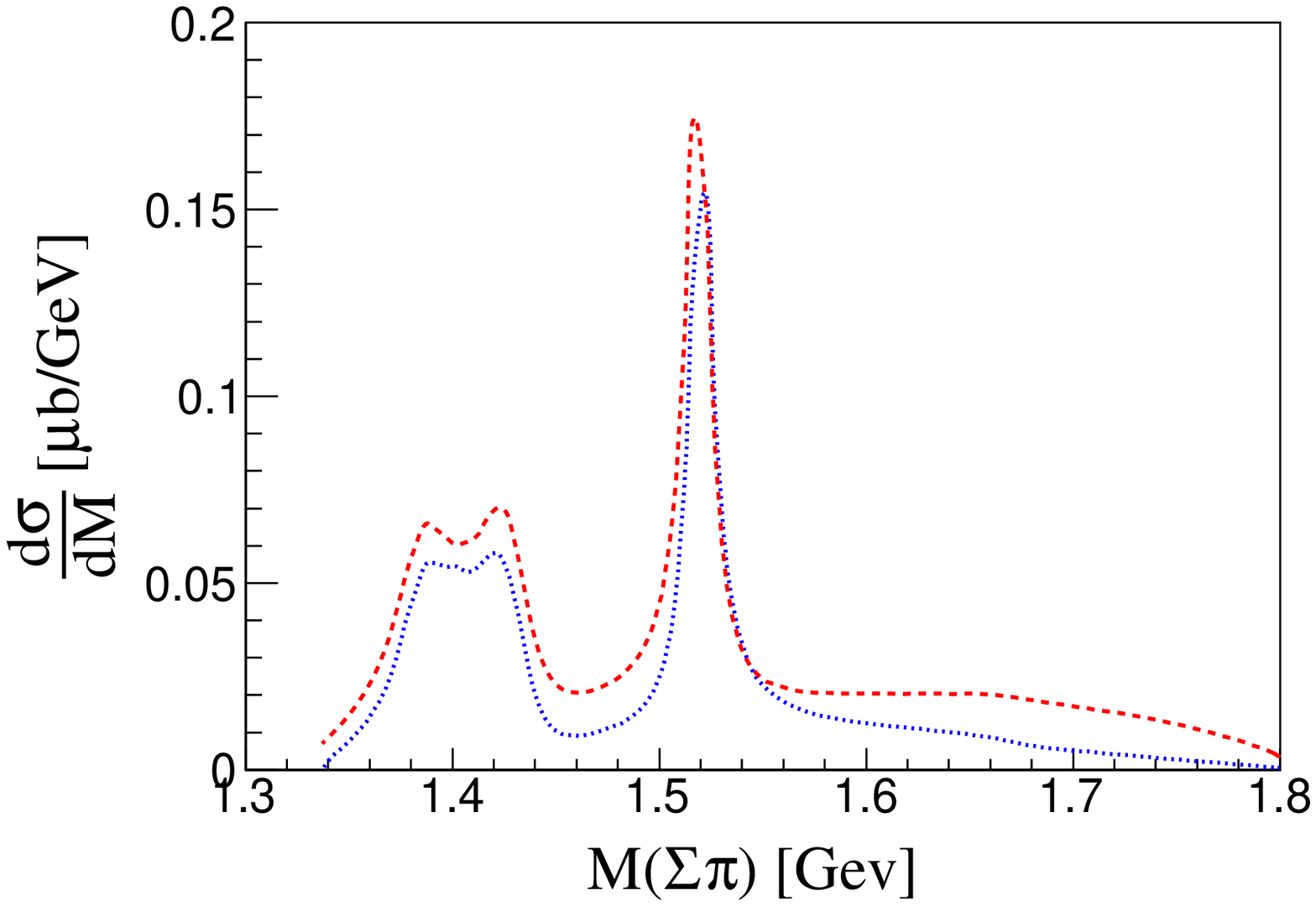}
    \caption{Model predictions \cite{Nam2010} to understand the difference in the $K^{*0}$ background contributions to the decay of $\Lambda^*$ into the two channels, $\Sigma^+\pi^-$ and $\Sigma^-\pi^+$, shown as short dashed and dotted curves, respectively. The predictions are shown as a function of $\Sigma\pi$ invariant mass, $M(\Sigma\pi)$.}
  \label{fig:dsigmadm}
\end{figure}

The difference in the strengths of the backgrounds, as seen in Fig.~\ref{fig:alllambdasigpsigm}, is caused by the unequal contributions from the $K^{*0}$ in the two measured $\Sigma\pi$ channels. Figure~\ref{fig:dsigmadm} shows calculations that include $K^{*0}$ background contributions to the cross sections of the two decay channels of the $\Lambda^*$ resonances. These calculations represented in the figure clarify that the intermediate $K^{*0} \rightarrow K^+ \pi^-$ decay is significantly responsible for introducing more background to the data in the $\Sigma^+\pi^-$ decay channel than in the $\Sigma^-\pi^+$ decay channel for the $\Lambda(1520)$. Therefore, these predictions provide an explanation to the contrasting backgrounds in the two channels, as seen in Fig.~\ref{fig:alllambdasigpsigm} and Fig.~\ref{fig:samplefitdata}. Very similar mass distributions of the $\Lambda^*$ resonances for the $\Sigma^{\pm}\pi^{\mp}$ decay channels can be seen in the study of the reaction $K^-p \rightarrow \Lambda(1520)\pi^{0}$ for the $\Sigma^{+}\pi^{-}\pi^{0}$ and $\Sigma^{-}\pi^{+}\pi^{0}$ final states by J. Griselin \textit{et al}.\ (1975) \cite{GRISELIN1975189}.

The $\Lambda(1520)$ events were selected by cutting on the $MM(K^+)$ distribution in the range from 1.44 to 1.60~GeV. This cut around the $\Lambda(1520)$ peak for the data is based on the particle data mass range for the $\Lambda(1520)$ and is consistent with the range of the $\Lambda(1520)$ peak obtained from our simulation of events for the two decay branches.

\subsubsection{Dalitz Plots}

\begin{figure}
   \includegraphics[width= 1\linewidth, keepaspectratio = true]{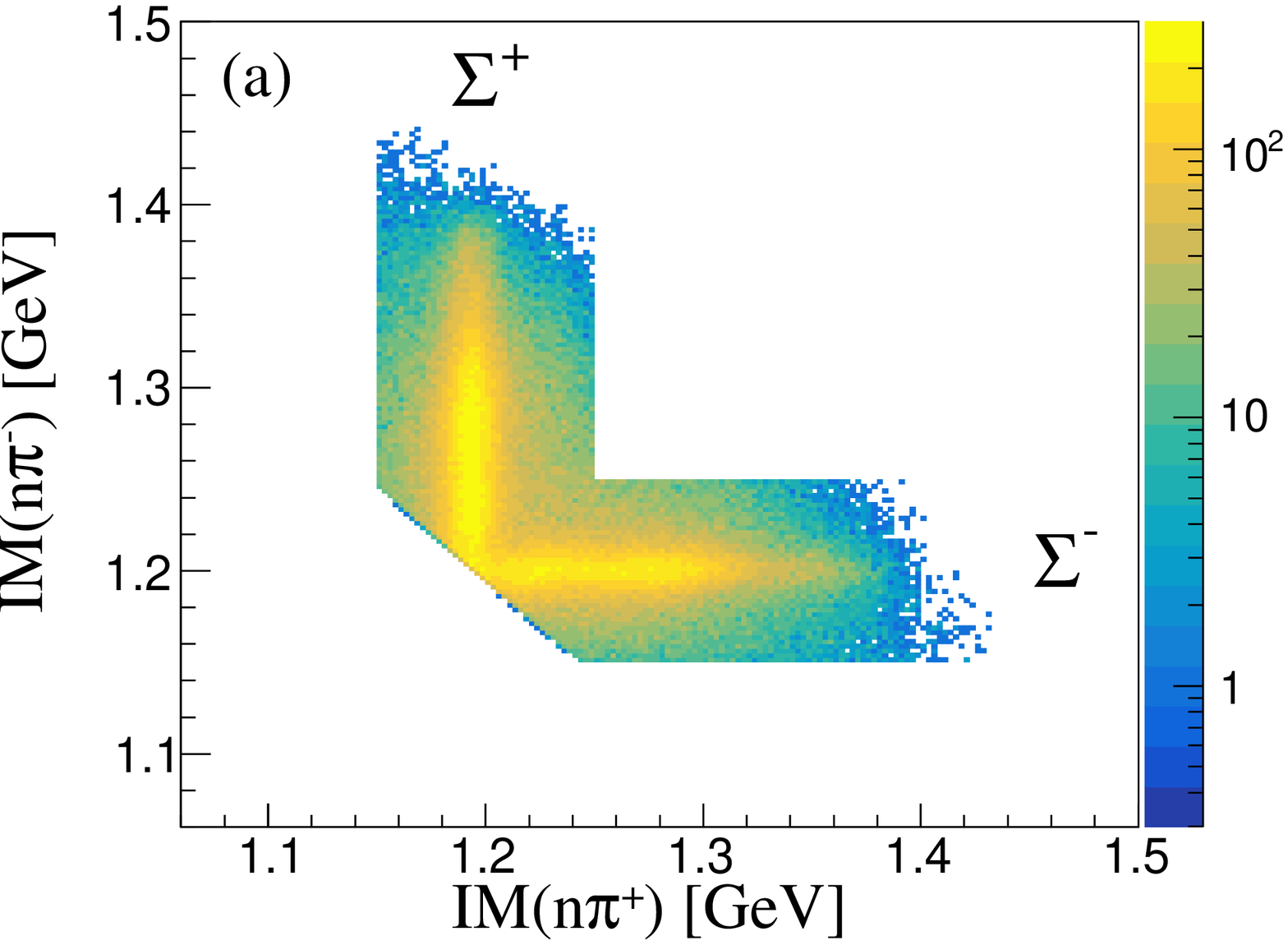}
   \includegraphics[width= 1\linewidth, keepaspectratio = true]{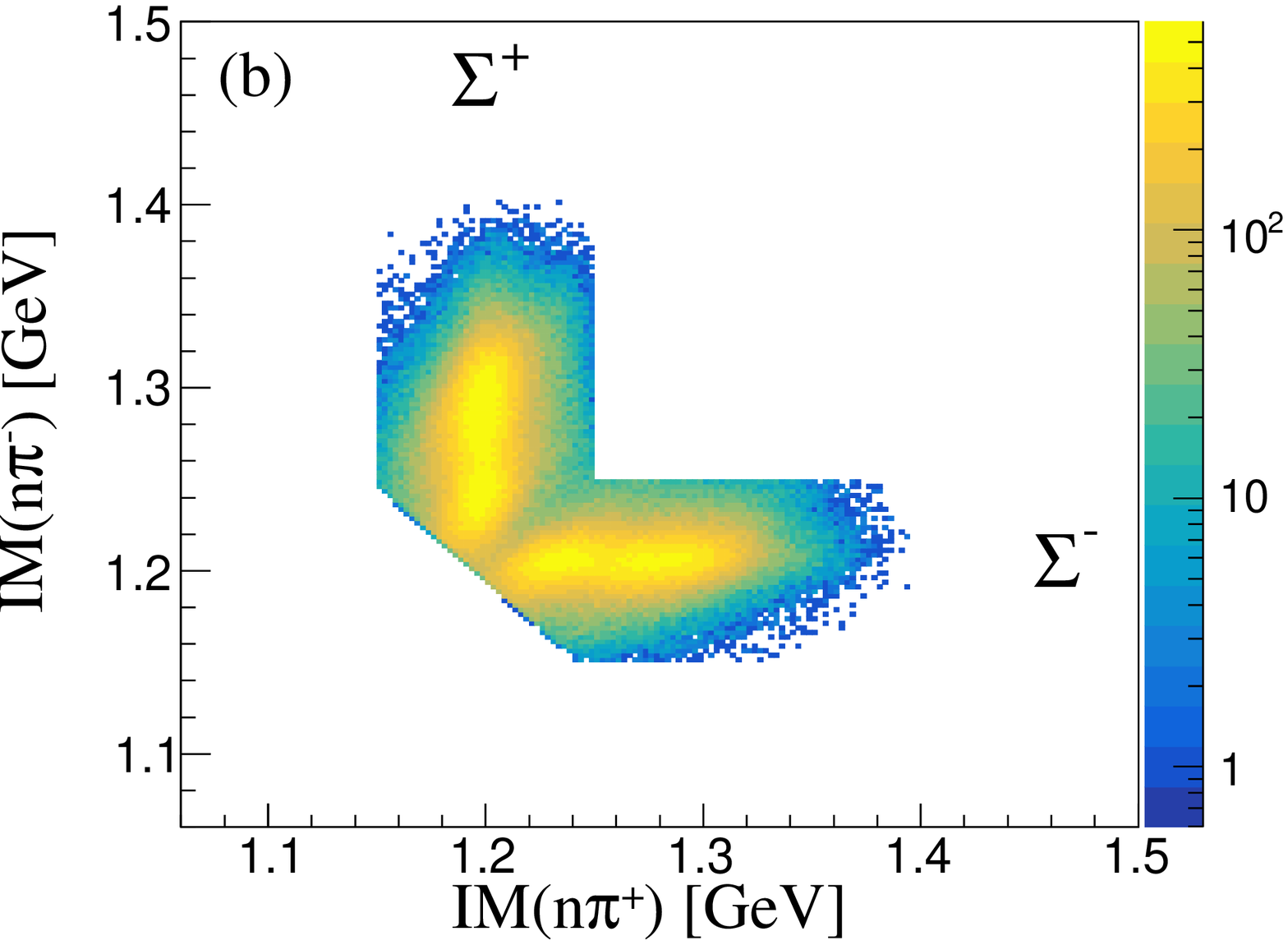}
    \caption{The $IM(n\pi^{-})$ vs. $IM(n\pi^{+})$ distribution for data in (a) and simulation in (b) with the vertical and horizontal strips reflecting $\Sigma^{+}$ and $\Sigma^{-}$, respectively are shown.}
  \label{fig:dalitzplots}
\end{figure}

The invariant mass ($IM$) distributions of different combinations of the final-state particles for our reaction were used with the goal of looking for physics backgrounds that could contribute to our final state. Plots of the decay $\Lambda(1520) \rightarrow n \pi^{+} \pi^{-}$ were studied by investigating the 2-D correlations of $IM(n\pi^{-})$ vs $IM(n\pi^{+})$ for both data and simulation as shown in Fig.~\ref{fig:dalitzplots}.

A remarkable similarity is seen between the data and the Monte-Carlo (MC) simulations. Figure~\ref{fig:dalitzplots} shows that there are clear bands corresponding to the $\Sigma^+$ and $\Sigma^-$ baryons, with very little overlap at the intersection. Events were assigned to only one branch, depending on whether the $IM$ was closer to the known mass of the $\Sigma^+$ or $\Sigma^-$ (for both data and MC).  Studies using the MC show that only about 1\% of events were misclassified, and the leakage was the same (within statistics) both ways.

The $IM(n\pi^{-})$ vs. $IM(n\pi^{+})$ plots for both the data and the simulation showed a region at the intersection that did not contribute to the $\Lambda(1520)$ peak. A diagonal cut was made to eliminate these events, which improved the signal-to-background ratio.

\subsection{\label{sec:simulation}Simulation}

A GEANT3-based Monte Carlo (MC) \cite{Brun1978} was used to simulate events for our experiment with the same final-state particles. Since, the acceptance of a detector is reaction dependent, the simulation for the two-decay channels of the $\Lambda(1520)$, the $\Sigma^+\pi^-$ and $\Sigma^-\pi^+$, was independently generated for the processes $\gamma p \rightarrow K^+ \Lambda(1520) \rightarrow K^+ \Sigma^+ \pi^- \rightarrow K^+ \pi^+ \pi^- (n)$ and $\gamma p \rightarrow K^+ \Lambda(1520) \rightarrow K^+ \Sigma^- \pi^+ \rightarrow K^+ \pi^+ \pi^- (n)$, respectively.

The MC event generator was based on the user input parameters and settings that include beam position, target material, reaction products, decay channels, and the $t$-slope parameter \cite{g12note}. The differential cross sections can be modeled as a function of $t$-slope by,
\begin{equation}
\label{tslopeFormula}
 \frac{d\sigma}{dt} = \sigma_0 e^{-bt},
\end{equation}
where $\frac{d\sigma}{dt}$ is the differential cross section, $b$ is the
$t$-slope parameter, and $\sigma_0$ is the amplitude of the cross section.
In order to best estimate the $b$-value for the simulation so that it matched our data, different values of $b$ were used to generate different sets of Monte-Carlo simulations, which were compared with the data distributions versus $t$. As a result, detector acceptance (or efficiency) was calculated with simulated events using $b = 1.5$~GeV$^{-2}$ and $b = 2.0$~GeV$^{-2}$ for $W\leq2.85$~GeV and $W>2.85$~GeV, respectively.

There were multiple triggers set up during the $g12$ experiment. The trigger relevant to our reaction is the one where events were recorded with three charged particles detected in three different sectors of CLAS. Due to the complex trigger configuration, the efficiency of the trigger was studied and accounted for by including it into the MC simulation. The same cuts, described previously for the data, were also applied for the simulations.

\subsection{Yield, Acceptance \& Luminosity}
\label{sec:yieldandacceptance}

\begin{figure*}
  \includegraphics[width= 0.49\textwidth, keepaspectratio = true]{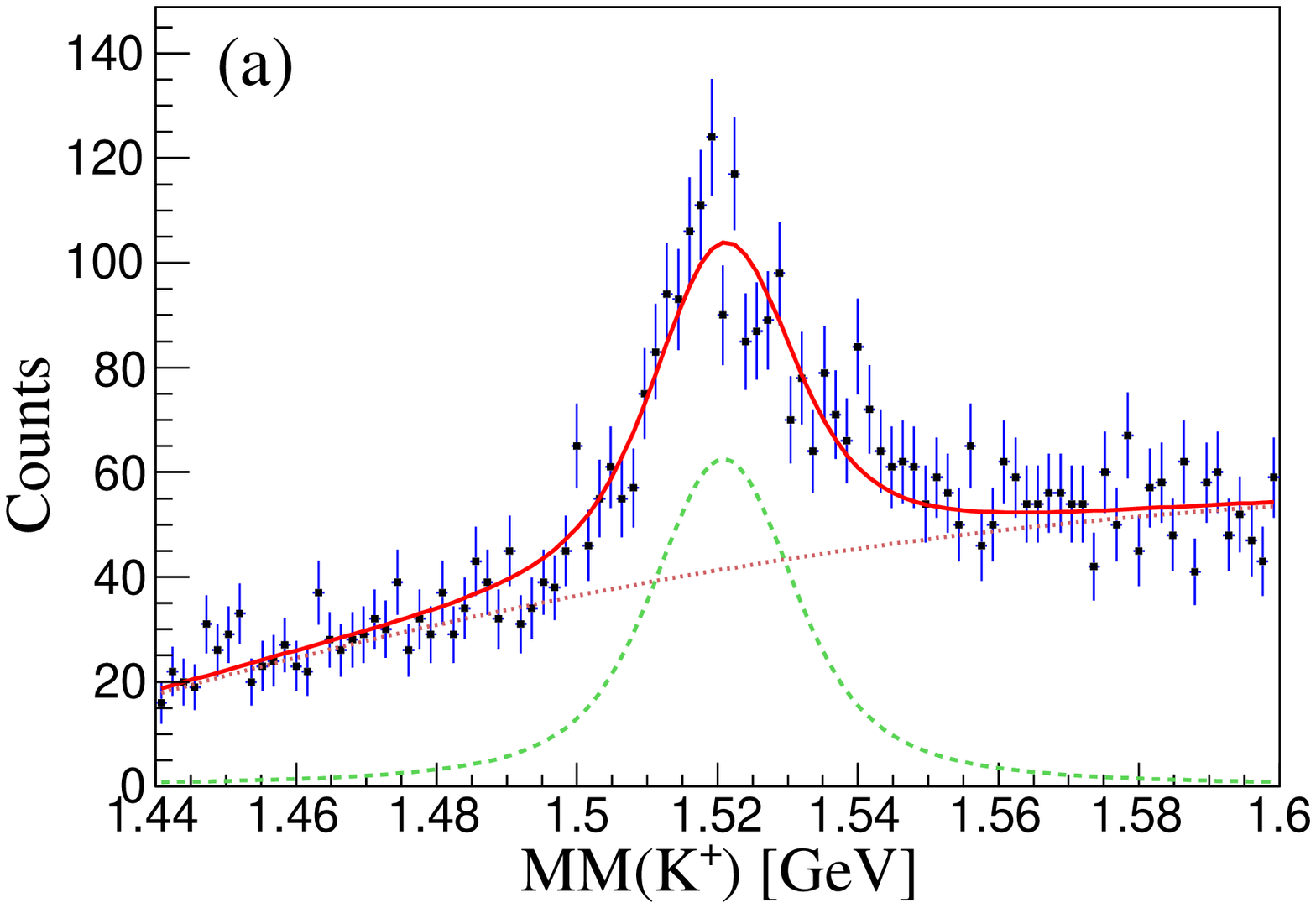}
  \includegraphics[width= 0.49\textwidth, keepaspectratio = true]{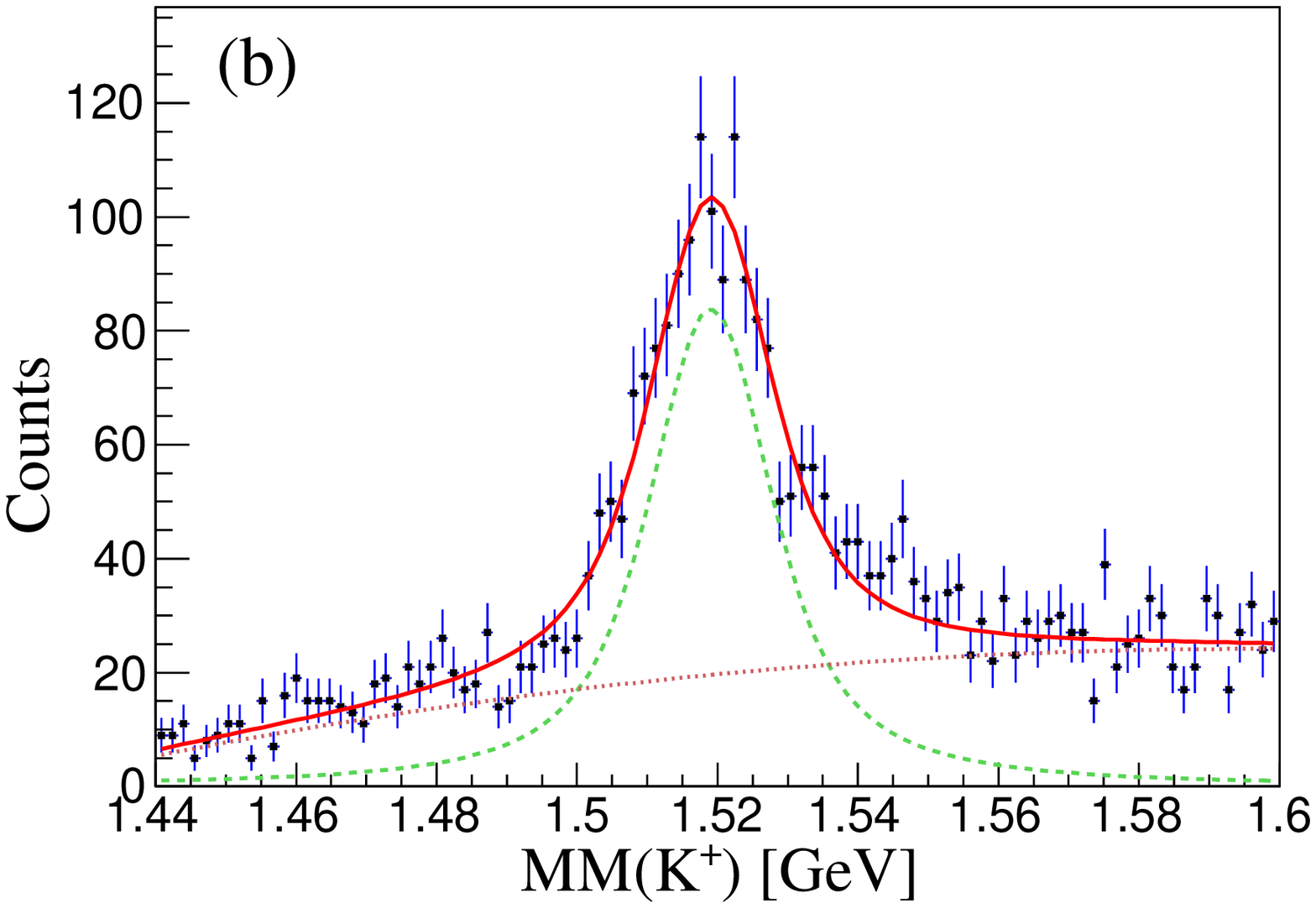}
\caption{The yield extract fits for the $\Lambda(1520)$ peak in the $MM(K^+)$ distribution for $2.25 < W$[GeV] $\leq 2.35$ and $0.5 \leq \cos{\theta}_{K^{+}}^{c.m.} \leq 0.7$ for data in the two decay channels $\Lambda(1520) \rightarrow \Sigma^+\pi^-$ and $\Lambda(1520) \rightarrow \Sigma^-\pi^+$, in (a) and (b), respectively are shown. The sum of the signal fit with a Voigtian function (dashed curve), along with the background estimated with a polynomial function (dotted curve), gives the total function (solid curve). The yield of the signal events is given by the integral of the signal fit curve.}
\label{fig:samplefitdata}
\end{figure*}

\subsubsection{Kinematic Binning}
\label{sec:binningscheme}
The events that made it through the event selection procedure, representing the $\Lambda(1520)$, were sorted into bins of center-of-mass (CM) energy ($W$) and azimuthal angle for the $K^+$ in the CM frame, $\cos{\theta}_{K^{+}}^{c.m.}$.
The CM energy, $W$, is a function of photon energy, $E_\gamma$, and the mass of the proton target, $m_p$. 

Events were also binned in center-of-mass (CM) energy, $W$, and squared four-momentum transfer $t$. The momentum transfer value is obtained from the Mandelstam $t$-variable for the reaction $\gamma p \rightarrow K^+ \Lambda(1520)$,
\begin{equation}
    t = (P_{\gamma}-P_{K^{+}})^2,
\end{equation}
where $P_{\gamma}$ and $P_{K^{+}}$ are the four-momenta of the incident photon and the detected $K^{+}$, respectively.

For the differential cross sections as a function of CM angle, 10 $W$-bins were taken in the range $2.25 < W$[GeV]$ \leq 3.25$, each of 100 MeV width. Each $W$ bin was studied with $\cos{\theta}_{K^{+}}^{c.m.}$ bins of width, $\Delta{\cos{\theta}_{K^{+}}^{c.m.}} = 0.2$, $-0.9 \leq \cos{\theta}_{K^{+}}^{c.m.} \leq 0.9$. For the cross sections as a function of $t$, nine $W$-bins were taken in the range $2.25 \leq W$[GeV]$ \leq 3.15$, each of 100 MeV width, where each $W$ bin was studied with various $t$ bins, $-2.5 \leq t$[GeV$^2$] $\leq -0.3$  with a width, $\Delta t = 0.2$ [GeV$^2$]. Due to bad photon Tagger scintillators, the events with $W$-values between 2.55~GeV and 2.6~GeV were omitted so that, the fourth $W$-bin has a 50~MeV width ($2.6 < W$[GeV]$ \leq 2.65$). 

\subsubsection{Yield Extraction}
\label{sec:yieldextraction}

The $\Lambda(1520)$ peak in the $MM(K^+)$ distribution was fit with a Gaussian-convoluted non-relativistic Breit-Wigner function, known as the Voigtian profile. A second-order polynomial function was chosen to estimate the smooth background. Figure~\ref{fig:samplefitdata} shows fitting samples for a particular kinematic bin. The Voigtian centroid parameter limits are set between 1.510 - 1.525~GeV, whereas the non-relativistic Breit-Wigner width is limited to 14.6 - 16.6~MeV. Both parameters are related to mass and full width values for the $\Lambda(1520)$ \cite{pdg2018,Qiang2010}. The background subtracted signal yield $Y(W,\cos{\theta}_{K^{+}}^{c.m.} \mbox{ or } t)$ was obtained by integrating the Voigtian function, as shown above by the region within the dashed curve in Fig.~\ref{fig:samplefitdata}.

\subsubsection{Acceptance}
\label{sec:acceptance}

The accepted number of events out of the total generated events, from the MC simulation provide a scale factor to correct the number of events in each kinematic bin $(W, \cos{\theta}_{K^{+}}^{c.m.} \mbox{ or } t)$. Hence, an acceptance value, $A(W, \cos{\theta}_{K^{+}}^{c.m.} \mbox{ or } t)$, was calculated as the ratio of the MC
events that were accepted to the total generated events and is given by
\begin{equation}
 A(W, \cos{\theta}_{K^{+}}^{c.m.} \mbox{ or } t) = \frac{Y_{acc}}{N_{gen}},
\end{equation}
where $Y_{acc}$ is the accepted yield of the simulated events and $N_{gen}$ 
is the total number of generated events.

The accepted events distributions were obtained using the MC accepted files for both the channels. These simulated files underwent a similar treatment to that of the data, including the cuts and corrections. Since the MC generates only the signal, the accepted event distribution was fit using a Voigtian function only. The non-relativistic Breit-Wigner width, $\sigma_L$, of the Voigtian function was kept fixed (for the MC fits only) at the physical width of the $\Lambda(1520)$, $\Gamma_{\Lambda(1520)} = 15.6$ MeV \cite{pdg2018}, for the fit.

\subsubsection{Luminosity}

The luminosity or flux, $L(W)$, was evaluated as,
\begin{equation}
 L(W) = \frac{\rho_p N_A l_t}{A_p}N_{\gamma}(W),
\end{equation}
where $N_{\gamma}(W)$ is the number of incident photons in a given $W$ range, $\rho_p = 0.07114$ g/cm$^3$ is the density of the proton target,
$l_t = 40$ cm is the target length, $N_A$ is Avogadro's number, and $A_p = 1.00794$ g/mol is the atomic mass of proton \cite{g12note}.

\section{Differential Cross Sections}
\label{sec:diffcross}

\begin{figure*}
    \centering
    \includegraphics[width=\linewidth, keepaspectratio = true]{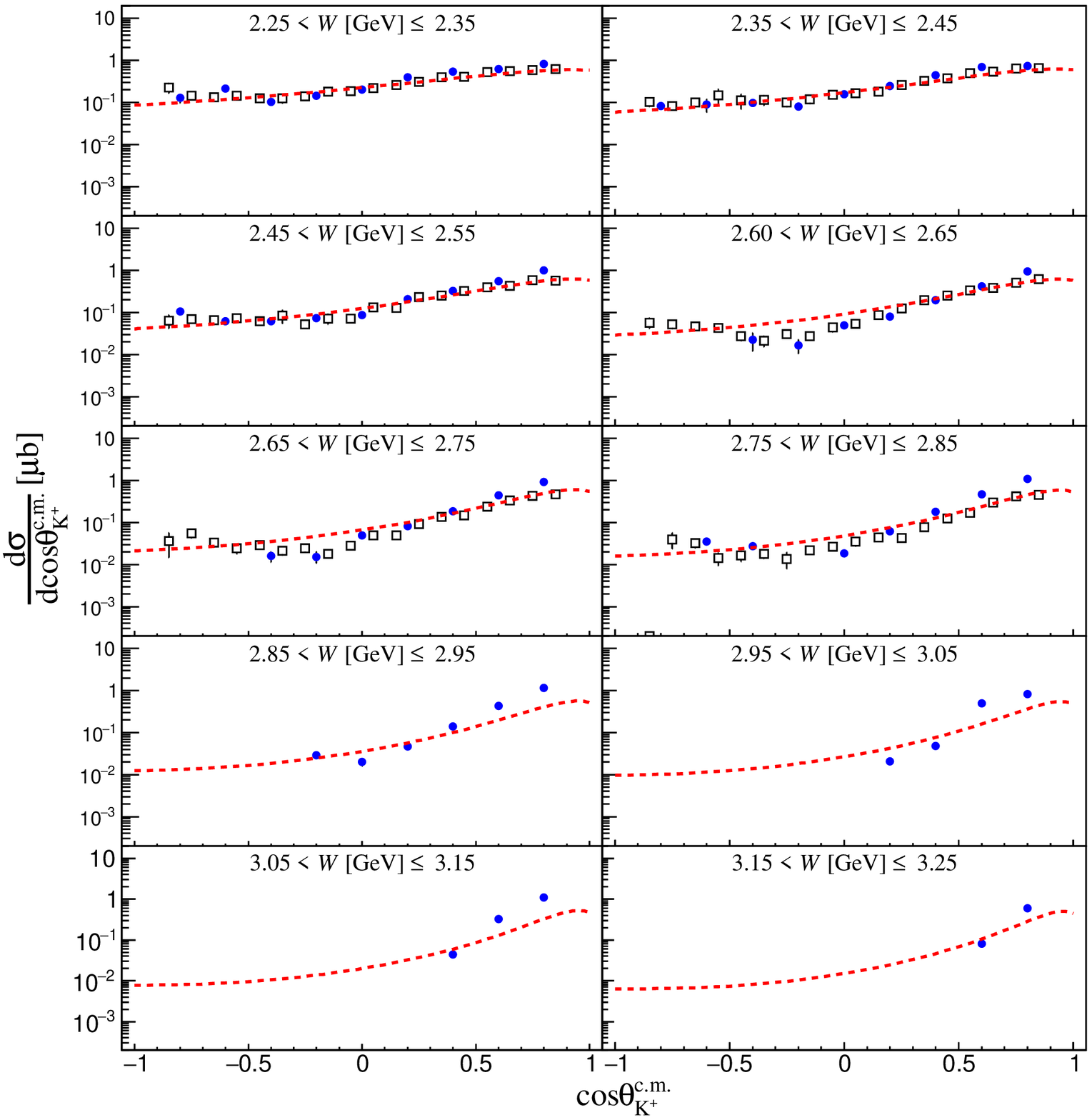}
    \caption{Differential cross sections for $\gamma p \rightarrow K^+ \Lambda(1520)$ are shown as solid circles for 10 center-of-mass energies ($W$) between $2.25 \leq W$~[GeV] $ \leq 3.25$ as a function of $\cos{\theta}_{K^{+}}^{c.m.}$. The hollow squares show the previous CLAS results by Moriya \textit{et al}.\ \cite{kmoriya}. The theoretical calculations from the model of \cite{Nam2010} are shown by the dashed curves. The error bars represent statistical uncertainties.}
    \label{fig:diffcross}
\end{figure*}
\begin{figure*}
    \centering
    \includegraphics[width=\linewidth, keepaspectratio = true]{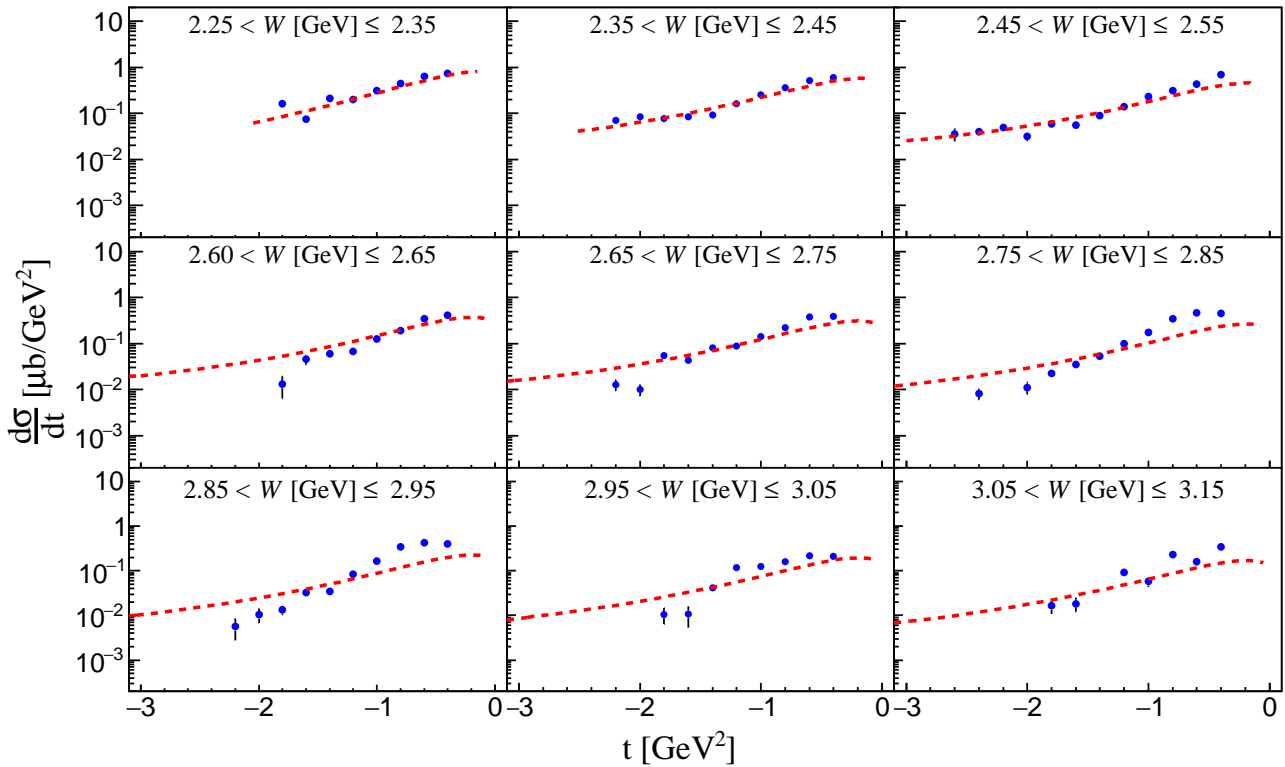}
    \caption{Differential cross sections for $\gamma p \rightarrow K^+ \Lambda(1520)$ are shown as solid circles for 9 center-of-mass energies ($W$) between $2.25 \leq W$~[GeV] $ \leq 3.15$ as a function of momentum transfer $t$. The theoretical calculations from the model of \cite{Nam2010} are shown by the dashed curves. The error bars represent statistical uncertainties.}
    \label{fig:tdiffcross}
\end{figure*}

The differential cross sections for the reaction $\gamma p \rightarrow K^+ \Lambda(1520)$ were calculated in $\cos{\theta}_{K^{+}}^{c.m.}$ bins using
\begin{equation}
 \frac{d\sigma}{d\cos{\theta}_{K^{+}}^{c.m.}} = \frac{Y(W,\cos{\theta}_{K^{+}}^{c.m.})}{\tau \Delta \cos{\theta}_{K^{+}}^{c.m.} A(W,\cos{\theta}_{K^{+}}^{c.m.}) L(W)} \times \gamma_{corr},
 \label{eqn:dcs}
\end{equation}
where $Y(W, \cos{\theta}_{K^{+}}^{c.m.})$ is the yield value and $A(W, \cos{\theta}_{K^{+}}^{c.m.})$ is the detector acceptance, $L(W)$ is the luminosity as a 
function of the center-of-mass ($W$) energy, and $\tau$ accounts for the branching ratio factors.

The differential cross sections were also obtained as a function of $t$ as,
\begin{equation}
 \frac{d\sigma}{dt} = \frac{Y(W,t)}{\tau \Delta t A(W,t) L(W)} \times \gamma_{corr},
 \label{eqn:dcst}
\end{equation}
where $Y(W, t)$ is the yield value and $A(W, t)$ is the detector acceptance. The kinematic bin widths, $\Delta\cos{\theta}_{K^{+}}^{c.m.} = 0.2$ and $\Delta t = 0.2$~GeV$^2$/$c^2$, represent the size of each
$\cos{\theta}_{K^{+}}^{c.m.}$-bin and $t$-bin, respectively. The photon multiplicity correction factor, $\gamma_{corr}$, is 1.03.

The decay modes of the $\Lambda(1520)$ include a branching ratio factor (b.r.) of 0.42 via the $\Sigma\pi$ channel into $\Sigma^{+}\pi^{-}$, $\Sigma^{-}\pi^{+}$ or $\Sigma^{0}\pi^{0}$. Using Clebsch-Gordon Coefficients, the individual b.r.~factor for the $\Sigma\pi$ channels comes out to be 0.14. The $\Sigma^{+}$ decays to a $n$ and $\pi^{+}$ with a b.r.~factor of 0.48, whereas the $\Sigma^{-}$ decays to a $n$ and a $\pi^{-}$ with a b.r.~of 1.0. Hence, the branching ratio used in Eqs.~\ref{eqn:dcs} and~\ref{eqn:dcst} for the $\Lambda(1520) \rightarrow \Sigma^+ \pi^-$ channel is 0.0672, whereas the branching ratio factor used for $\Lambda(1520) \rightarrow \Sigma^- \pi^+$ channel is 0.14. The branching ratio factors, $\tau$, were applied separately to each decay channel in order to obtain the $\Lambda(1520)$ differential cross sections for each decay mode. The differential cross sections for the two branches are then averaged to obtain the $\Lambda(1520)$ differential cross sections. The uncertainties were determined by standard propagation of errors.

The differential cross sections as a function of $\cos{\theta}_{K^{+}}^{c.m.}$ are shown in Fig.~\ref{fig:diffcross}. The figure shows the current analysis in comparison with previous CLAS results \cite{kmoriya} (hollow squares). The results by Moriya \textit{et al}.~\cite{kmoriya}, uses a $W$ bin width of 100 MeV and $\cos{\theta}_{K^{+}}^{c.m.}$ bin width of 0.1. It can be seen that there is good agreement between the differential cross sections for the $\Lambda(1520)$ between this analysis and previous CLAS results. The theoretical calculations provided by Seung-il Nam are also shown (as the dashed curves) in the figure.

Similarly, the differential cross sections as a function of $t$ are shown in Fig.~\ref{fig:tdiffcross}. The theoretical calculations provided by Seung-il Nam are also shown (as the dashed curves) in the figure.

\section{Systematic Uncertainties}
\label{sec:sysunc}

\begin{table}[b]
\caption{Summary of the systematic uncertainties calculated in this analysis.}
\label{tab:totsys}
\begin{ruledtabular}
\begin{tabular}{lr}
 \textbf{Source}                                            & \textbf{Uncertainty} \\ \hline
$t$-slope dependence           &    0.78\%             \\
Timing cut                     &    4.11\%             \\
Minimum $|p|$ cut              &    0.20\%             \\ 
$z$-vertex cut                 &    1.28\%             \\
Fiducial cut                   &    3.13\%             \\
Background function  &    2.07\%             \\
Signal integral range           &    0.43\%           \\
Flux consistency/luminosity \cite{g12note}   &    5.70\%             \\
Sector-by-sector \cite{g12note}               &    5.90\%             \\
Target \cite{g12note}            &    0.50\%             \\
\hline
\textbf{Total Systematic Uncertainty} &   \textbf{10.05\%}  \\
\end{tabular}
\end{ruledtabular}
\end{table}

The systematic uncertainties for this study were estimated by making variations on the different cuts and taking the
average relative difference in the final result. Hence, the systematic uncertainty is understood
as the shift of the average of the relative differences from zero; zero being no net change in the result
after a variation. The variation in the parameters was done by observing the data and making an estimate of what range is a reasonable choice for each systematic uncertainty. For instance, a first-order polynomial function was taken as a variation to the nominal choice of second-order polynomial function for estimating the background. Similarly, tighter fiducial boundaries on the active region of the detector were used as a deviation from the normal cut on those boundaries to determine the uncertainty due to the fiducial cut. Similar variations are used for the other parameters.

The systematic uncertainties for this analysis, both general and specific, are summarized in Table~\ref{tab:totsys}. The general systematic uncertainties that refer to the uncertainties due to the $g12$ run conditions, for instances, flux consistency/luminosity, sector-by-sector, and target, are outlined in \cite{g12note}. For instance, the sector-by-sector uncertainty is computed by the deviation of the acceptance-corrected yields in each sector of the CLAS detector \cite{clas} from the average acceptance-corrected yield of all six sectors. Similarly, the uncertainty from the target accounts for the variations of the pressure and temperature throughout the $g12$ data-taking period. The reaction specific uncertainties that depend on the analysis process, cuts, and corrections performed during the study are also reported. Each of the systematic effects has its contribution to the total systematic uncertainty of this analysis. The total systematic uncertainty of 10.05\% is calculated by the sum in quadrature of all systematic uncertainties, assuming that they are independent of each other.

\section{\label{sec:discussion}Discussion and Conclusions}
As seen in Figs.~\ref{fig:diffcross} and~\ref{fig:tdiffcross}, the model calculations by Nam are in good agreement with our experimental results. The theory calculations, represented by dashed curves, are the numerical results without the $N^*$ contribution, and conserve gauge invariance \cite{Nam2010}. The calculations with the $N^*$ contribution (not shown in Figs~\ref{fig:diffcross} and~\ref{fig:tdiffcross}) indicate that the $N^*$ contribution in the $s$-channel process is very small \cite{Nam2010}, and only slightly changes the calculation in the first $W$ bin. This is because only $N^*$ resonances with mass below 2.2~GeV are included in such calculations. For the $u$-channel, where there is intermediate $\Lambda^*$ exchange, such an approach is not needed to explain the experimental data as it does not significantly contribute to the calculation due to the small coupling constant of the proton-hyperon vertex.

Consequently, even for the higher-energy region up to $W = 3.2$~GeV, the present simple Born model guided by gauge invariance, which is represented by the inclusion of the contact-term contribution and appropriate form factor prescription to conserve gauge invariance in terms of the Rarita-Schwinger formalism, can describe the data qualitatively well. In this sense, from a theoretical point of view, gauge invariance is a powerful guide to understand the electromagnetic (EM) coupling of the production of spin-3/2 baryons. The introduction of the contact term simply conserves gauge invariance in the photon-nucleon infinitesimal point interaction scenario.

The theory calculations modeled with a $K$-exchange diagram, as shown in Fig.~\ref{fig:reaction}, qualitatively reproduce the data without Regge, $K^*$, and hyperon resonances. Hence, we can conclude that the simplest theoretical model with a pseudoscalar $K$-meson exchange, assuming $t$-channel dominance, is sufficient to explain the broad features of our data, without the need for the inclusion of other reaction processes. 

The slight increases above the theory observed in the data in the backward scattering region could be improved by the inclusion of hyperon resonances in the $u$-channel, although the theoretical uncertainties, such as the EM transition couplings between the hyperons, are considerable for the $u$-channel. Also, some small deviations of the model compared to the data at forward angles for $W > 2.7$~GeV, may need more sophisticated theoretical approaches. One can think that these differences could be explained by higher-spin $N^*$ resonances and the Regge trajectories. Although not shown here, a more detailed theoretical study \cite{Nam2010} has concluded that the $K^*$-$N$-$\Lambda(1520)$ coupling must be very small in order to reproduce the data.

Even though the theoretical calculations are adequate to explain the broad behavior of the experimental cross sections, we can say that there is an indication that at higher $W$ there is a possibility for $K^*$ exchange or a possible interference of a $K^*$-exchange with the $K$-exchange. These future improvements may introduce a small correction to the theoretical predictions. Hence, this study can contribute to a better understanding of the $\Lambda(1520)$ using higher-energy photoproduction data.

Although our results do not show any evidence for higher-mass $N^*$ resonances decaying to the $K^+ \Lambda(1520)$ final state, the lack of such evidence is useful in itself. One question that has surrounded the “missing resonances” problem is whether an $N^*$ could have a strong preference to decay into strangeness channels. For example, there is some evidence from photoproduction of $K^{*+} \Lambda$ that a few higher-mass $N^*$ states have a significant decay branch to that final state \cite{Anisovich2017}. The present results indicate that these same higher-mass $N^*$ states, if confirmed, do not contribute to the $K^+ \Lambda(1520)$ final state. The lack of $N^*$ states contributing in the $s$-channel to the current cross sections is a constraint on the branching ratios of possible higher-mass $N^*$ states. Further exploration of the ``missing resonances'' at higher mass would be better done using either the $K^* \Lambda$ or the $\gamma p \rightarrow \pi^+ \pi^- p$ reaction \cite{mokeev2020}. A systematic study of the latter reaction, using virtual photons, is being carried out with the CLAS12 detector \cite{CLAS12} (an upgrade of the CLAS detector) at Jefferson Lab and also at other facilities.

We acknowledge the staff of the Accelerator and Physics Divisions at the Thomas Jefferson National Accelerator Facility who made this experiment possible. This work was supported in part by 
the Chilean Comisi\'on Nacional de Investigaci\'on Cient\'ifica y Tecnol\'ogica (CONICYT),
the Italian Istituto Nazionale di Fisica Nucleare,
the French Centre National de la Recherche Scientifique,
the French Commissariat \`{a} l'Energie Atomique,
the U.S. Department of Energy,
the National Science Foundation,
the Scottish Universities Physics Alliance (SUPA),
the United Kingdom's Science and Technology Facilities Council,
and the National Research Foundation of Korea. The Southeastern Universities Research Association (SURA) operates the
Thomas Jefferson National Accelerator Facility for the United States
Department of Energy under contract DE-AC05-06OR23177.

\appendix
\section{Data Table}
\begin{longtable*}[h]{ccc}
\caption{Differential cross sections for $\gamma p \rightarrow K^{+} \Lambda(1520)$, as a function of CM angle. The uncertainties represent only the statistical contributions.\label{tab:dcsL1520data}}\\
\hline\hline
$W$~[GeV] & $\cos{\theta}_{K^{+}}^{c.m.}$ & ${d\sigma}/{d\cos{\theta}_{K^{+}}^{c.m.}}$~[$\mu$b] \\
\hline
\endfirsthead
Table~\ref{tab:dcsL1520data} \textit{Continued..}\\
\hline\hline
$W$~[GeV] & $\cos{\theta}_{K^{+}}^{c.m.}$ & ${d\sigma}/{d\cos{\theta}_{K^{+}}^{c.m.}}$~[$\mu$b] \\
\hline
\endhead
\hline\hline
\endfoot
(2.25, 2.35) & ($-$0.9, $-$0.7) & 0.128 $\pm$ 0.031 \\
(2.25, 2.35) & ($-$0.7, $-$0.5) & 0.211 $\pm$ 0.021 \\
(2.25, 2.35) & ($-$0.5, $-$0.3) & 0.103 $\pm$ 0.014 \\
(2.25, 2.35) & ($-$0.3, $-$0.1) & 0.145 $\pm$ 0.012 \\
(2.25, 2.35) & ($-$0.1, 0.1) & 0.203 $\pm$ 0.013 \\
(2.25, 2.35) & (0.1, 0.3) & 0.398 $\pm$ 0.020 \\
(2.25, 2.35) & (0.3, 0.5) & 0.534 $\pm$ 0.022 \\
(2.25, 2.35) & (0.5, 0.7) & 0.618 $\pm$ 0.024 \\
(2.25, 2.35) & (0.7, 0.9) & 0.833 $\pm$ 0.033 \\
(2.35, 2.45) & ($-$0.9, $-$0.7) & 0.081 $\pm$ 0.015 \\
(2.35, 2.45) & ($-$0.7, $-$0.5) & 0.089 $\pm$ 0.030 \\
(2.35, 2.45) & ($-$0.5, $-$0.3) & 0.096 $\pm$ 0.009 \\
(2.35, 2.45) & ($-$0.3, $-$0.1) & 0.079 $\pm$ 0.008 \\
(2.35, 2.45) & ($-$0.1, 0.1) & 0.156 $\pm$ 0.010 \\
(2.35, 2.45) & (0.1, 0.3) & 0.247 $\pm$ 0.013 \\
(2.35, 2.45) & (0.3, 0.5) & 0.443 $\pm$ 0.019 \\
(2.35, 2.45) & (0.5, 0.7) & 0.693 $\pm$ 0.027 \\
(2.35, 2.45) & (0.7, 0.9) & 0.739 $\pm$ 0.030 \\
(2.45, 2.55) & ($-$0.9, $-$0.7) & 0.106 $\pm$ 0.020 \\
(2.45, 2.55) & ($-$0.7, $-$0.5) & 0.062 $\pm$ 0.011 \\
(2.45, 2.55) & ($-$0.5, $-$0.3) & 0.061 $\pm$ 0.008 \\
(2.45, 2.55) & ($-$0.3, $-$0.1) & 0.074 $\pm$ 0.008 \\
(2.45, 2.55) & ($-$0.1, 0.1) & 0.087 $\pm$ 0.007 \\
(2.45, 2.55) & (0.1, 0.3) & 0.208 $\pm$ 0.012 \\
(2.45, 2.55) & (0.3, 0.5) & 0.327 $\pm$ 0.013 \\
(2.45, 2.55) & (0.5, 0.7) & 0.562 $\pm$ 0.020 \\
(2.45, 2.55) & (0.7, 0.9) & 1.015 $\pm$ 0.037 \\
(2.60, 2.65) & ($-$0.5, $-$0.3) & 0.022 $\pm$ 0.010 \\
(2.60, 2.65) & ($-$0.3, $-$0.1) & 0.017 $\pm$ 0.006 \\
(2.60, 2.65) & ($-$0.1, 0.1) & 0.050 $\pm$ 0.009 \\
(2.60, 2.65) & (0.1, 0.3) & 0.079 $\pm$ 0.009 \\
(2.60, 2.65) & (0.3, 0.5) & 0.199 $\pm$ 0.016 \\
(2.60, 2.65) & (0.5, 0.7) & 0.417 $\pm$ 0.023 \\
(2.60, 2.65) & (0.7, 0.9) & 0.941 $\pm$ 0.049 \\
(2.65, 2.75) & ($-$0.5, $-$0.3) & 0.016 $\pm$ 0.004 \\
(2.65, 2.75) & ($-$0.3, $-$0.1) & 0.015 $\pm$ 0.005 \\
(2.65, 2.75) & ($-$0.1, 0.1) & 0.049 $\pm$ 0.005 \\
(2.65, 2.75) & (0.1, 0.3) & 0.081 $\pm$ 0.008 \\
(2.65, 2.75) & (0.3, 0.5) & 0.185 $\pm$ 0.011 \\
(2.65, 2.75) & (0.5, 0.7) & 0.445 $\pm$ 0.013 \\
(2.65, 2.75) & (0.7, 0.9) & 0.934 $\pm$ 0.038 \\
(2.75, 2.85) & ($-$0.7, $-$0.5) & 0.036 $\pm$ 0.007 \\
(2.75, 2.85) & ($-$0.5, $-$0.3) & 0.027 $\pm$ 0.004 \\
(2.75, 2.85) & ($-$0.1, 0.1) & 0.019 $\pm$ 0.003 \\
(2.75, 2.85) & (0.1, 0.3) & 0.062 $\pm$ 0.006 \\
(2.75, 2.85) & (0.3, 0.5) & 0.181 $\pm$ 0.010 \\
(2.75, 2.85) & (0.5, 0.7) & 0.464 $\pm$ 0.020 \\
(2.75, 2.85) & (0.7, 0.9) & 1.103 $\pm$ 0.043 \\
(2.85, 2.95) & ($-$0.3, $-$0.1) & 0.029 $\pm$ 0.005 \\
(2.85, 2.95) & ($-$0.1, 0.1) & 0.020 $\pm$ 0.004 \\
(2.85, 2.95) & (0.1, 0.3) & 0.047 $\pm$ 0.006 \\
(2.85, 2.95) & (0.3, 0.5) & 0.139 $\pm$ 0.009 \\
(2.85, 2.95) & (0.5, 0.7) & 0.436 $\pm$ 0.021 \\
(2.85, 2.95) & (0.7, 0.9) & 1.171 $\pm$ 0.049 \\
(2.95, 3.05) & (0.1, 0.3) & 0.021 $\pm$ 0.004 \\
(2.95, 3.05) & (0.3, 0.5) & 0.048 $\pm$ 0.008 \\
(2.95, 3.05) & (0.5, 0.7) & 0.494 $\pm$ 0.025 \\
(2.95, 3.05) & (0.7, 0.9) & 0.830 $\pm$ 0.049 \\
(3.05, 3.15) & (0.3, 0.5) & 0.044 $\pm$ 0.007 \\
(3.05, 3.15) & (0.5, 0.7) & 0.326 $\pm$ 0.022 \\
(3.05, 3.15) & (0.7, 0.9) & 1.086 $\pm$ 0.056 \\
(3.15, 3.25) & (0.5, 0.7) & 0.081 $\pm$ 0.011 \\
(3.15, 3.25) & (0.7, 0.9) & 0.592 $\pm$ 0.055 \\
\end{longtable*}

\begin{longtable*}[h]{ccc}
\caption{Differential cross sections for $\gamma p \rightarrow K^{+} \Lambda(1520)$ as a function of $t$. The uncertainties represent only the statistical contributions.\label{tab:dcsL1520datat}}\\
\hline\hline
$W$~[GeV] & $t$~[GeV$^2$] & ${d\sigma}/{dt}$~[$\mu$b/GeV$^2$] \\
\hline
\endfirsthead
Table~\ref{tab:dcsL1520datat} \textit{Continued..}\\
\hline\hline
$W$~[GeV] & $t$~[GeV$^2$] & ${d\sigma}/{dt}$~[$\mu$b/GeV$^2$] \\
\hline
\endhead
\hline\hline
\endfoot
(2.25, 2.35) & ($-$1.9, $-$1.7) & 0.164 $\pm$ 0.025 \\
(2.25, 2.35) & ($-$1.7, $-$1.5) & 0.076 $\pm$ 0.011 \\
(2.25, 2.35) & ($-$1.5, $-$1.3) & 0.213 $\pm$ 0.016 \\
(2.25, 2.35) & ($-$1.3, $-$1.1) & 0.198 $\pm$ 0.015 \\
(2.25, 2.35) & ($-$1.1, $-$0.9) & 0.314 $\pm$ 0.017 \\
(2.25, 2.35) & ($-$0.9, $-$0.7) & 0.448 $\pm$ 0.019 \\
(2.25, 2.35) & ($-$0.7, $-$0.5) & 0.640 $\pm$ 0.023 \\
(2.25, 2.35) & ($-$0.5, $-$0.3) & 0.735 $\pm$ 0.028 \\
(2.35, 2.45) & ($-$2.3, $-$2.1) & 0.070 $\pm$ 0.011 \\
(2.35, 2.45) & ($-$2.1, $-$1.9) & 0.084 $\pm$ 0.010 \\
(2.35, 2.45) & ($-$1.9, $-$1.7) & 0.077 $\pm$ 0.010 \\
(2.35, 2.45) & ($-$1.7, $-$1.5) & 0.084 $\pm$ 0.010 \\
(2.35, 2.45) & ($-$1.5, $-$1.3) & 0.091 $\pm$ 0.009 \\
(2.35, 2.45) & ($-$1.3, $-$1.1) & 0.161 $\pm$ 0.010 \\
(2.35, 2.45) & ($-$1.1, $-$0.9) & 0.251 $\pm$ 0.013 \\
(2.35, 2.45) & ($-$0.9, $-$0.7) & 0.359 $\pm$ 0.016 \\
(2.35, 2.45) & ($-$0.7, $-$0.5) & 0.524 $\pm$ 0.022 \\
(2.35, 2.45) & ($-$0.5, $-$0.3) & 0.594 $\pm$ 0.025 \\
(2.45, 2.55) & ($-$2.7, $-$2.5) & 0.036 $\pm$ 0.011 \\
(2.45, 2.55) & ($-$2.5, $-$2.3) & 0.040 $\pm$ 0.007 \\
(2.45, 2.55) & ($-$2.3, $-$2.1) & 0.049 $\pm$ 0.006 \\
(2.45, 2.55) & ($-$2.1, $-$1.9) & 0.032 $\pm$ 0.006 \\
(2.45, 2.55) & ($-$1.9, $-$1.7) & 0.060 $\pm$ 0.007 \\
(2.45, 2.55) & ($-$1.7, $-$1.5) & 0.055 $\pm$ 0.007 \\
(2.45, 2.55) & ($-$1.5, $-$1.3) & 0.091 $\pm$ 0.007 \\
(2.45, 2.55) & ($-$1.3, $-$1.1) & 0.140 $\pm$ 0.008 \\
(2.45, 2.55) & ($-$1.1, $-$0.9) & 0.232 $\pm$ 0.011 \\
(2.45, 2.55) & ($-$0.9, $-$0.7) & 0.316 $\pm$ 0.013 \\
(2.45, 2.55) & ($-$0.7, $-$0.5) & 0.430 $\pm$ 0.017 \\
(2.45, 2.55) & ($-$0.5, $-$0.3) & 0.705 $\pm$ 0.030 \\
(2.60, 2.65) & ($-$1.9, $-$1.7) & 0.013 $\pm$ 0.007 \\
(2.60, 2.65) & ($-$1.7, $-$1.5) & 0.046 $\pm$ 0.011 \\
(2.60, 2.65) & ($-$1.5, $-$1.3) & 0.060 $\pm$ 0.009 \\
(2.60, 2.65) & ($-$1.3, $-$1.1) & 0.067 $\pm$ 0.008 \\
(2.60, 2.65) & ($-$1.1, $-$0.9) & 0.125 $\pm$ 0.013 \\
(2.60, 2.65) & ($-$0.9, $-$0.7) & 0.194 $\pm$ 0.017 \\
(2.60, 2.65) & ($-$0.7, $-$0.5) & 0.342 $\pm$ 0.024 \\
(2.60, 2.65) & ($-$0.5, $-$0.3) & 0.417 $\pm$ 0.033 \\
(2.65, 2.75) & ($-$2.3, $-$2.1) & 0.013 $\pm$ 0.003 \\
(2.65, 2.75) & ($-$2.1, $-$1.9) & 0.010 $\pm$ 0.003 \\
(2.65, 2.75) & ($-$1.9, $-$1.7) & 0.055 $\pm$ 0.006 \\
(2.65, 2.75) & ($-$1.7, $-$1.5) & 0.043 $\pm$ 0.005 \\
(2.65, 2.75) & ($-$1.5, $-$1.3) & 0.080 $\pm$ 0.006 \\
(2.65, 2.75) & ($-$1.3, $-$1.1) & 0.089 $\pm$ 0.008 \\
(2.65, 2.75) & ($-$1.1, $-$0.9) & 0.140 $\pm$ 0.009 \\
(2.65, 2.75) & ($-$0.9, $-$0.7) & 0.220 $\pm$ 0.012 \\
(2.65, 2.75) & ($-$0.7, $-$0.5) & 0.377 $\pm$ 0.021 \\
(2.65, 2.75) & ($-$0.5, $-$0.3) & 0.386 $\pm$ 0.024 \\
(2.75, 2.85) & ($-$2.3, $-$2.1) & 0.008 $\pm$ 0.004 \\
(2.75, 2.85) & ($-$2.1, $-$1.9) & 0.011 $\pm$ 0.003 \\
(2.75, 2.85) & ($-$1.9, $-$1.7) & 0.023 $\pm$ 0.003 \\
(2.75, 2.85) & ($-$1.7, $-$1.5) & 0.035 $\pm$ 0.004 \\
(2.75, 2.85) & ($-$1.5, $-$1.3) & 0.054 $\pm$ 0.005 \\
(2.75, 2.85) & ($-$1.3, $-$1.1) & 0.098 $\pm$ 0.008 \\
(2.75, 2.85) & ($-$1.1, $-$0.9) & 0.175 $\pm$ 0.011 \\
(2.75, 2.85) & ($-$0.9, $-$0.7) & 0.341 $\pm$ 0.017 \\
(2.75, 2.85) & ($-$0.7, $-$0.5) & 0.473 $\pm$ 0.023 \\
(2.75, 2.85) & ($-$0.5, $-$0.3) & 0.450 $\pm$ 0.027 \\
(2.85, 2.95) & ($-$2.3, $-$2.1) & 0.006 $\pm$ 0.003 \\
(2.85, 2.95) & ($-$2.1, $-$1.9) & 0.010 $\pm$ 0.004 \\
(2.85, 2.95) & ($-$1.9, $-$1.7) & 0.013 $\pm$ 0.003 \\
(2.85, 2.95) & ($-$1.7, $-$1.5) & 0.033 $\pm$ 0.005 \\
(2.85, 2.95) & ($-$1.5, $-$1.3) & 0.034 $\pm$ 0.003 \\
(2.85, 2.95) & ($-$1.3, $-$1.1) & 0.083 $\pm$ 0.008 \\
(2.85, 2.95) & ($-$1.1, $-$0.9) & 0.164 $\pm$ 0.012 \\
(2.85, 2.95) & ($-$0.9, $-$0.7) & 0.342 $\pm$ 0.019 \\
(2.85, 2.95) & ($-$0.7, $-$0.5) & 0.421 $\pm$ 0.022 \\
(2.85, 2.95) & ($-$0.5, $-$0.3) & 0.398 $\pm$ 0.034 \\
(2.95, 3.05) & ($-$1.9, $-$1.7) & 0.010 $\pm$ 0.004 \\
(2.95, 3.05) & ($-$1.7, $-$1.5) & 0.011 $\pm$ 0.005 \\
(2.95, 3.05) & ($-$1.5, $-$1.3) & 0.042 $\pm$ 0.006 \\
(2.95, 3.05) & ($-$1.3, $-$1.1) & 0.119 $\pm$ 0.010 \\
(2.95, 3.05) & ($-$1.1, $-$0.9) & 0.125 $\pm$ 0.016 \\
(2.95, 3.05) & ($-$0.9, $-$0.7) & 0.157 $\pm$ 0.018 \\
(2.95, 3.05) & ($-$0.7, $-$0.5) & 0.219 $\pm$ 0.023 \\
(2.95, 3.05) & ($-$0.5, $-$0.3) & 0.209 $\pm$ 0.030 \\
(3.05, 3.15) & ($-$1.7, $-$1.5) & 0.018 $\pm$ 0.006 \\
(3.05, 3.15) & ($-$1.5, $-$1.3) & 0.009 $\pm$ 0.002 \\
(3.05, 3.15) & ($-$1.3, $-$1.1) & 0.091 $\pm$ 0.013 \\
(3.05, 3.15) & ($-$1.1, $-$0.9) & 0.057 $\pm$ 0.013 \\
(3.05, 3.15) & ($-$0.9, $-$0.7) & 0.228 $\pm$ 0.019 \\
(3.05, 3.15) & ($-$0.7, $-$0.5) & 0.157 $\pm$ 0.017 \\
(3.05, 3.15) & ($-$0.5, $-$0.3) & 0.345 $\pm$ 0.038 \\
\end{longtable*}

\bibliography{L1520ref.bib}
\end{document}